# Microstructure Thermal Stability and Superplastic Behavior of Al-6%Mg-0.12%Sc-0.10%Zr-0.10%(Yb, Er, Hf) Ultrafine-Grained Alloys

V.N. Chuvil'deev [a], M.Yu. Gryaznov [a], S.V. Shotin [a], A.V. Nokhrin [a,1], G.S. Nagicheva [a,b], C.V. Likhnitskii [a], I.S. Shadrina [a], V.I. Kopylov [a], A.A. Bobrov [a], M.K. Chegurov [a], O.E. Pirozhnikova [a]

[a] Lobachevsky State University of Nizhny Novgorod, Nizhny Novgorod, Russia

[b] Lakehead University, Thunder Bay, Canada

nokhrin@nifti.unn.ru

**Abstract**

Superplastic behavior of ultrafine-grained (UFG) Al-6%Mg-0.12%Sc-0.10%Zr-0.1%X alloys, where X = Yb (Alloy #1-Yb), Er (Alloy #2-Er), and Hf (Alloy #3-Hf), has been studied. The total content of Sc, Zr, Yb, Er, Hf in the alloys was 0.32 wt.% (0.117–0.118 at.%). The alloys used for benchmarking were Al-6%Mg-0.12%Sc-0.20%Zr (Alloy #4-Zr) and Al-6%-0.22%Sc-0.10%Zr (Alloy #5-Sc). Their UFG microstructure was formed with Equal Channel Angular Pressing. Two different types of deformation behavior during superplasticity were demonstrated. A simultaneous increase in yield stress and elongation to failure during superplastic deformation was discovered. High deformation temperatures were shown to cause a competition between dynamic (strain-induced) grain growth and dynamic recrystallization, leading to a finer grain microstructure. The values of strain hardening factor ($n$), strain rate sensitivity factor ($m$), and superplastic deformation threshold stress ($\sigma_p$) were determined. The impact of the type and concentration of alloying elements on the deformation behavior and dynamic grain growth of UFG Al-6%Mg alloys was investigated. It was established that the maximum elongation to failure in Alloy #1-Yb and Alloy #2-Er is observed at lower deformation temperatures than in Alloy #4-Zr and Alloy #5-Sc. The superplastic properties of Alloy #3-Hf are superior to those of Alloy #4-Zr and Alloy #5-Sc with high content of alloying elements (in at.%). Alloy #1-Yb manifests good elongation to failure ($\delta$ = 910%) at low temperatures

---

[1] Corresponding author (nokhrin@nifti.unn.ru)

(400 ºC). The satisfiability of Hart's criterion for calculating uniform deformation value under superplastic conditions was verified. It was demonstrated that cavitation when pores are formed in large Al$_3$X particles at high temperatures causes early failure of aluminum alloys.

**Abbreviations:** AEs – Alloying Elements; ECAP – Equal Channel Angular Pressing; EDS – Energy Dispersion (analysis); GBS – Grain Boundary Sliding; RT – Room Temperature; SEM – Scanning Electron Microscopy; SER – Specific electrical resistivity; UFG (alloy) – ultrafine-grained (alloy).

**1. Introduction**

Industrial Al-6%Mg aluminum alloys with Sc and/or Zr additives (grades 1570, 1570C, 1571, 1575, 1580, 1597, etc.) have good strength and corrosive resistance at room temperature (RT), are easy to weld, and have high impact strength at RT and low temperatures [1]. Additional deformation treatment of Al-6%Mg-(Sc, Zr) alloys helps to form a ultrafine-grained (UFG) microstructure, to increase the strength of the alloys [2], their fatigue strength [3], and to ensure high superplastic properties at elevated temperatures [4, 5]. The combination of these properties makes Al-6%Mg-(Sc, Zr) fine-grained alloys a highly appealing material for manufacturing new high-strength corrosion-resistant products for aircraft and shipbuilding industries.

High superplastic properties of aluminum alloys (high elongation to failure, high uniform deformation, low yield stress, high optimum strain rate, low optimum deformation temperature) are necessary for effective superplastic sheet pressing of complex-shaped products. One way to ensure these properties efficiently is to reduce the average grain size to the submicron level [4-6]. Hence, ensuring the stability of a non-equilibrium microstructure of UFG aluminum alloys should be prioritized. This is normally achieved by adding alloying elements (AE) to aluminum alloys. Scandium, when heated, helps to form Al$_3$Sc nanoparticles, making it one of the most efficient alloying elements [7, 8]. Al$_3$Sc coherent particles lend high thermal stability to UFG aluminum [9, 10]. Magnesium reduces grain-boundary diffusivity in aluminum, further helping to boost thermal stability of a non-equilibrium UFG microstructure [11].

Researchers are currently trying to replace expensive scandium with cheaper rare-earth elements (REEs) or transition metals (TMs). Additives of Yb, Er, Hf, and some others AE [12-14] which, when heated, form Al$_3$X particles with the L1$_2$ [15, 16] structure, are being considered as a substitute for scandium. It should be noted that Yb, Er, Hf can be an effective replacement for zirconium in Al-Mg-(Sc,Zr) alloys, as they can accelerate precipitation of Al$_3$X particles at temperatures that are lower than those required to form Al$_3$(Sc$_x$Zr$_{1-x}$) particles [17]. This makes it possible to stabilize a non-equilibrium UFG microstructure in highly deformed aluminum alloys more efficiently. An increased thermal stability of the microstructure can have a favorable impact on superplastic properties of UFG aluminum alloys.

The purpose of this research is to study (i) the efficiency of replacing 0.1 wt.% Sc with 0.1 wt.% Yb, Er, Hf at a constant zirconium content (0.1 wt.% Zr) in an Al-6%Mg-Sc-Zr alloy; (ii) the efficiency of replacing 0.1 wt.% Zr with 0.1% Yb, Er, Hf in relation to the deformation behavior of the Al-6%Mg-Sc-Zr alloy at a constant scandium content (0.12 wt.% Sc). The main focus of our research is to study microstructure thermal stability during annealing, superplastic deformation behavior, and fracture under superplasticity of Al-6%Mg-Sc-Zr UFG alloys.

**2. Materials and Methods**

The study focused on Al-6 wt.%Mg-0.12wt.%Sc-0.10wt.%Zr alloys with 0.1 wt. % Yb (Alloy #1-Yb), Er (Alloy #2-Er), and Hf additions (Alloy #3-Hf). Alloys #1-Yb, #2-Er, #3-Hf had a total content of Sc, Zr, Yb, Er, Hf of 0.32 wt.% or 0.117-0.118 at.%. They were benchmarked against Al-6 wt.%Mg-0.12wt.%Sc-0.20wt.%Zr alloy (Alloy #4-Zr) and Al-6 wt.% Mg-0.22 wt.%Sc-0.10 wt.%Zr alloy (Alloy #5-Sc). The chemical composition of the alloys under study is presented in Table 1. The properties of Alloy #4-Zr were compared with those of Alloys #1–3 through analyzing the impact of replacing 0.1 wt.% Zr with 0.1% Yb, Er, Hf on the alloy superplastic behavior at a constant scandium content (0.12 wt.% Sc). The properties of Alloys #1–3 were compared with those of Alloy #5-Sc through analyzing the efficiency of substituting 0.1 wt.% Sc with 0.1 wt. % Yb, Er, Hf at a constant zirconium content (0.10 wt.% Zr).

Table 1 – Chemical composition of the aluminum alloys

| No Alloy | Alloying elements, wt.% / at.% [1] | | | | | | | |
|---|---|---|---|---|---|---|---|---|
| | Al | Mg | Sc | Zr | Yb | Er | Hf | Σ [2] |
| 1-Yb | Balance | 6.0 | 0.12 | 0.10 | 0.10 | - | - | 0.32 |
| | | 6.7 | 0.072 | 0.03 | 0.016 | | | 0.118 |
| 2-Er | | 6.0 | 0.12 | 0.10 | - | 0.10 | - | 0.32 |
| | | 6.7 | 0.072 | 0.03 | | 0.016 | | 0.118 |
| 3-Hf | | 6.0 | 0.12 | 0.10 | - | - | 0.10 | 0.32 |
| | | 6.7 | 0.072 | 0.03 | | | 0.015 | 0.117 |
| 4-Zr | | 6.0 | 0.12 | 0.20 | - | - | - | 0.32 |
| | | 6.7 | 0.072 | 0.06 | | | | 0.132 |
| 5-Sc | | 6.0 | 0.22 | 0.10 | - | - | - | 0.32 |
| | | 6.7 | 0.13 | 0.03 | | | | 0.160 |

[1] Numerator – compositions in wt.%, denominator – compositions in at.%.
[2] Total content of Sc, Zr, Yb, Er, Hf

Aluminum alloy were obtained by vacuum induction casting. An Indutherm VTC 200V Vacuum Casting Machine was used to produce ingots that were cast in a copper mold of 22×22×160 mm. The alloys were made from A99 grade aluminum, Mg90 grade magnesium, and Al-10%Zr, Al-2%Sc, Al-3%Yb, Al-3%Hf, Al-3%Er master alloys. Master alloys were obtained by induction casting followed by rolling into a foil 0.2 mm thick.

The UFG microstructure was formed with Equal Channel Angular Pressing (ECAP). The 22×22×160 mm workpieces were machine-worked to remove the inhomogeneous surface layer before ECAP (see the findings of cast alloys studies below). After milling, the samples with a 20×20 mm cross-section were ECAPed. The samples were not subjected to additional heat treatment prior to ECAP. The above workpieces were subjected to 3 cycles of ECAP at 275 ºC to form a UFG microstructure. ECAP was performed using a Ficep HF400L hydraulic press. Pressing was arranged using square-section tooling, with a channel intersection angle of 90°. ECAP was performed in line with B mode and a strain rate of 0.4 mm/s. After every ECAP cycle, the workpiece was cooled down to RT, fitted into the working channel, greased, mounted inside the working channel of an ECAP punch, and heated for 15 min before the next ECAP cycle. Graphite grease laced with molybdenum disulfide $MoS_2$ was used for ECAP.

The resulting flat specimens shaped as a dogbone-shaped with a working part of 2×2×3 mm were subjected to mechanical tests. Tensile tests were conducted using a Tinius Olsen H25K-S machine at a strain rate $\dot{\varepsilon}$ of $3.3 \cdot 10^{-3}$ s$^{-1}$ (tensile rate of $10^{-2}$ mm/s). Tests were run at RT and at temperatures ranging from 300 to 500 ºC. Additionally, at 400, 450, and 500 °C, the effect of strain rate $\dot{\varepsilon}$ ($3.3 \cdot 10^{-3}$, $3.3 \cdot 10^{-2}$, $3.3 \cdot 10^{-1}$ s$^{-1}$) on superplasticity characteristics in UFG alloys was studied. Heating the sample to the test temperature did not exceed 10 min. While testing, stress ($\sigma$)–strain ($\varepsilon$) curve was being registered and used to determine ultimate tensile strength ($\sigma_b$) and elongation to failure ($\delta$), as well uniform strain ($\varepsilon^*$) corresponding to ultimate tensile strength. The superplasticity test results were analyzed using the following equation [18, 19]:

$$\dot{\varepsilon} = A(\sigma^*/G)^{1/m}(b/d)^p(D_{eff}/b^2)(G\Omega/kT), \qquad (1)$$

where $m$ is the strain rate sensitivity coefficient of flow stress, $p$ is a numerical parameter equal to 2 or 3, $b$ is the Burgers vector, $G$ is the shear modulus, $k$ is the Boltzmann constant, $\sigma^* = \sigma_b - \sigma_p$ is the flow stress, $\sigma_p$ - threshold stress, and $D_{eff} = D_0 \exp(Q_{eff}/kT)$ is the effective diffusion coefficient under superplasticity, where $Q_{eff}$ is the apparent activation energy of superplastic flow.

Microstructure of the alloys was studied with a Leica DM IRM metallographic microscope and Jeol JSM-6490 scanning electron microscope (SEM) with Oxford Instruments INCA-350 EDS detector. The fractographic analysis that followed tensile tests was performed using a Jeol JSM-6490 SEM. Specimens were preliminarily subjected to mechanical grinding and polishing, as well as electropolishing (3 A, 30 V) for 1 min in $CrO_3 + H_3PO_4$ electrolyte. An alloy microstructure was exposed by etching in a solution of HF (15 mL) + $HNO_3$ (10 mL) + glycerol (35 mL). Microhardness (HV) was measured using a HVS-1000 hardness tester under a 50g load. The average accuracy of Hv measurement was ±25 MPa. Microstructure studies and microhardness measurements were carried out in the fracture area (Zone I) and in the undeformed area (Zone II) of the specimen. Specific electrical resistivity (SER) was measured through eddy current method with a SIGMATEST 2.069 instrument, using an 8 mm probe. The uncertainty of measuring the SER was ±0.03 µΩ·cm.

The cast and UFG aluminum alloy samples were annealed in an EKPS-10 air furnace. The accuracy of temperature control was ±10 ºC.

## 3. Results

### 3.1. Alloy characterizations

Figure 1 shows photographs of a cross-section of the central part of the cast alloy samples. The alloys are numbered as per Table 1. Alloy #4-Zr with high zirconium content and Alloy #1-Yb manifest a uniform macrostructure; the entire section is covered with small equiaxed grains 50–150 µm in size. All other alloys (#2-Er, #3-Hf, #5-Sc) are characterized by elongated dendritic grains along the edges of the sample and a uniform fine-grained microstructure in the cross-section of the central part. This result implies that a solid solution was partially decomposed during crystallization and primary $Al_3X$ particles were formed. As per [17], primary particles can modify the structure of cast alloys and, as a result, the structure of ingots is refined. No large pores or other casting flaws can be detected in the central portion of the ingot. There are β-phase particles along grain boundaries, which are easily destroyed when the sample is etched (Fig. 1e).

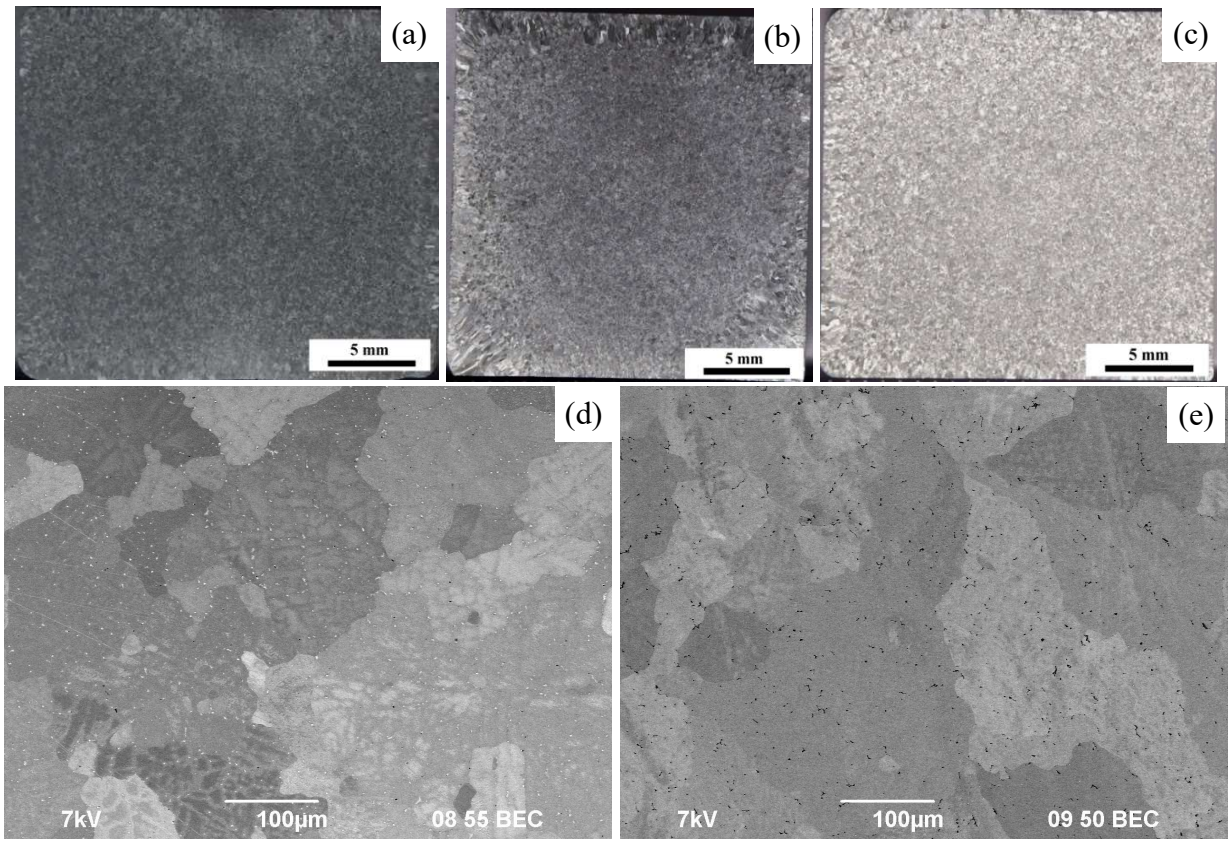

Figure 1. Macrostructure of cast Alloys #1-Yb (a, d), #3-Hf (b, e), #4-Zr (c)

There are single primary $Al_3Zr$ particles in cast Alloy #4-Zr, up to several μm in size. There are primary $Al_3Yb$ and $Al_3Er$ particles in cast Alloys #1-Yb and #2-Er, respectively, containing a large amount of magnesium (Fig. 2). The size of primary particles varies from 1 to 3 μm, primary particles being evenly distributed throughout the volume of the material (Fig. 1d). Isolated primary particles can be found in Alloys #3-Hf, #4-Zr, and #5-Sc, but their number is insignificant.

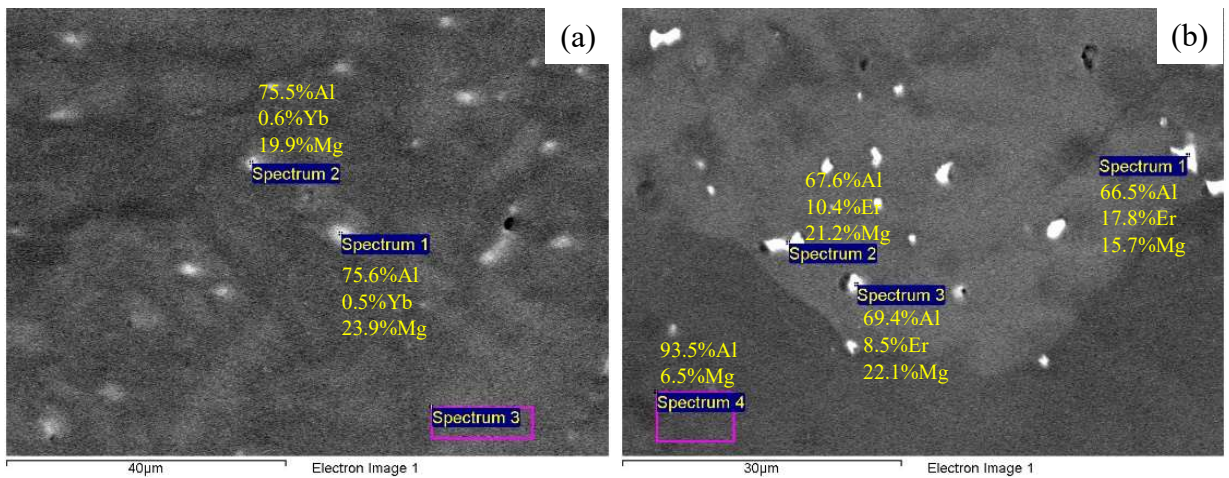

Figure 2. EDS analysis of primary particles composition in Alloys #1-Yb (a), #2-Er (b). SEM

A homogeneous fine-grained microstructure with an average grain size of ~0.5 μm is formed in the alloys following ECAP. Post-ECAP primary particles have size and shape similar to those of primary particles in cast alloys. No precipitation of secondary $Al_3X$ (X = Sc, Zr, Er, Hf, Yb) particles was detected after ECAP.

Table 2 shows the properties of cast alloys in the initial state (before annealing). A heterogeneous macrostructure causes a noticeable scatter in SER values in the cast alloys cross section. As can be seen from Table 2, the SER of the central part is less than at the edges of the samples. Metal solidifies slower in the center than at the edges; as a result, the amount of AEs transferred from the melt to the solid solution of the solidified metal will differ in the cross section of the samples. Differences in the concentration of AEs in the solid solution in the center and at the edges of the sample bring about the SER difference of ~0.1–0.2 μΩ·cm, the maximum SER difference between the center and the edges of the sample being 0.22 ± 0.02 μΩ·cm (Alloy #5-Sc). The range of the effect detected (~0.1–0.2 μΩ·cm) is approximately the same for all Al-6%Mg-Sc-Zr-(Yb,Er,Hf) alloys. This suggests that the effect observed is primarily due to heterogeneous magnesium distribution. The contribution of magnesium to the aluminum SER is known to be 0.49 μΩ·cm/at.% [20]. Therefore, the SER difference observed (~0.1–0.2 μΩ·cm) corresponds to a difference in magnesium concentration of ~0.17–0.36 wt.%Mg. According to GOST 4784-2019 (ISO 209:2007, EN 573-3:2013), the allowable deviation in magnesium concentration for AMg6 alloy is between 5.8 and 6.8 wt.%, for alloy 01570 – between 5.3 and 6.3 wt.% Mg, for alloy 01570C – between 5.0 and 5.6 wt.% Mg, for alloy 1575 – between 5.4 and 6.4 wt.% Mg. Hence, the observed scatter in magnesium concentrations (~0.17–0.36 wt.%Mg) in the test samples meets the requirements of the standard.

Table 2. Cast and UFG alloys characterization

| | Alloy #1-Yb | | Alloy #2-Er | | Alloy #3-Hf | | Alloy #4-Zr | | Alloy #5-Sc | |
|---|---|---|---|---|---|---|---|---|---|---|
| | Cast | UFG | Cast | UFG | Cast | UFG | Cast | UFG | Cast | UFG |
| Hv, MPa | 775 ± 85 | 1095 ± 55 | 775 ± 60 | 1060 ± 75 | 795 ± 70 | 1025 ± 50 | 800 ± 30 | 1090 ± 85 | 780 ± 80 | 1125 ± 55 |
| $\sigma_y$, MPa | 260 ± 10 | 375 ± 15 | 305 ± 15 | 325 ± 20 | 300 ± 15 | 375 ± 20 | 300 ± 15 | 370 ± 25 | 290 ± 20 | 340 ± 20 |
| $\rho_{exp}$, $\mu\Omega\cdot cm$ | 6.01-6.21(*) | 6.05-6.27(*) | 6.01-6.20(*) | 6.12-6.35(*) | 6.23-6.38(*) | 6.05-6.26(*) | 6.23-6.37(*) | 6.09-6.28(*) | 6.24-6.46(*) | 6.23-6.38(*) |
| $\rho_{th}$, $\mu\Omega\cdot cm$ | 6.42 | | 6.41 | | 6.33 | | 6.49 | | 6.53 | |

Note: (*) The first (smaller) number is the SER of the central part of the sample with a 22×22 mm cross-section with a uniform microstructure, the second (bigger) number is the SER at the edge of the sample where elongated dendritic grains are observed (see Fig. 1).

There is no noticeable change in the SER of the aluminum alloys following ECAP (Table 2). It can be assumed that a rise in the SER caused by increased flaw density during ECAP is offset by partial decomposition of a Al-(Sc,Zr) solid solution during warm ECAP. Post-ECAP UFG alloys have higher hardness than cast alloys (Table 2). The hardness of UFG alloys is higher than that of cast alloys by 250–350 MPa. The chemical composition of alloys does not have a noticeable impact on the hardness of unannealed cast and UFG alloys (Table 2).

Figure 3 demonstrates the dependences of hardness and SER on the temperature of 30-minute annealing of cast (Fig. 3a) and UFG alloys (Fig. 3b). Figure 3 shows SER values for the central part of UFG samples.

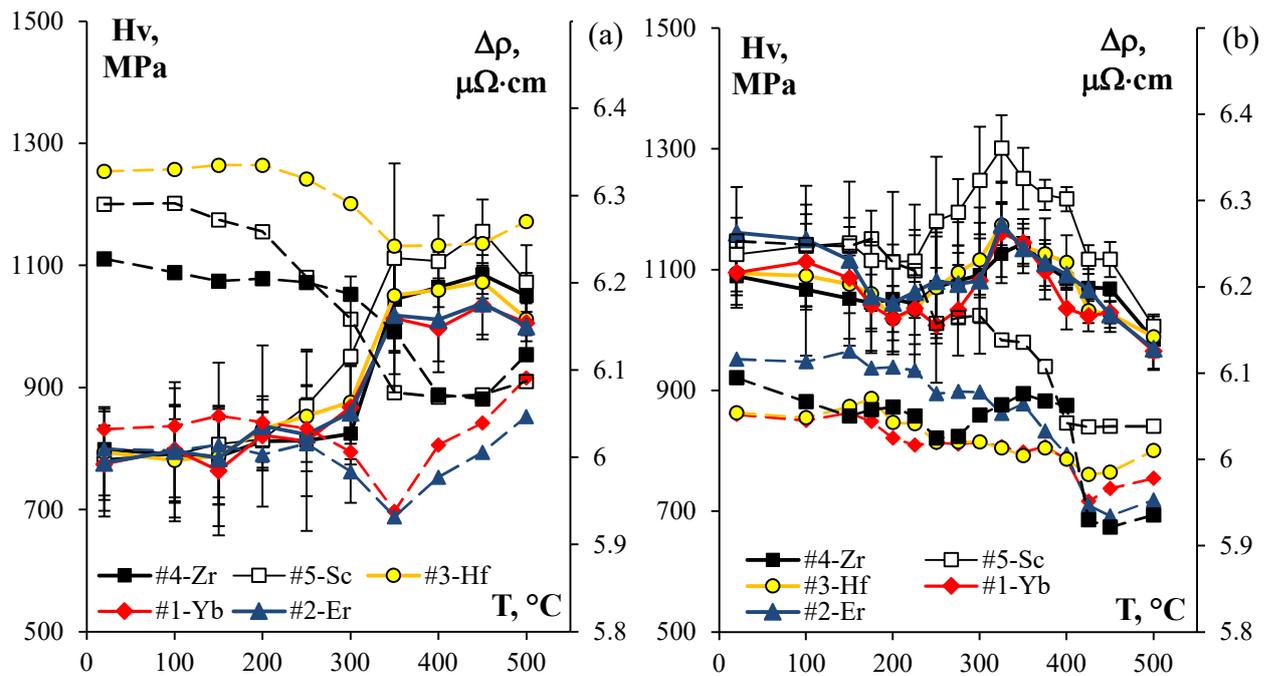

Figure 3. Dependences of microhardness (Hv) and SER changes (Δρ) on the temperature of 30-minute annealing of cast alloys (a) and UFG alloys (b). Solid lines indicate Hv(T) dependences while dotted lines show Δρ(T) dependences.

Figure 3a suggests that the largest change in the SER and hardness of cast alloys is observed for Alloy #5-Sc. Significant increases in Δρ and ΔHv are also observed for Alloy #4-Zr. The variation in Δρ for cast alloys additionally alloyed with Yb, Er, Hf is less significant than for Alloys #5-Sc and #4-Zr. A reduced SER at annealing temperatures over 400 °C is driven by partial dissolution of β-phase particles. This leads to an increased concentration of Mg in the crystal lattice and a decreased SER of the aluminum alloy.

Fig. 3b shows the findings of UFG alloys studies. It can be observed from Fig. 3b that Alloy #5-Sc manifests the biggest change in hardness ΔHv and SER Δρ. Secondary particles precipitate most actively in Alloy #2-Er, its properties (ΔHv, Δρ) being close to those of Alloy #4-Zr. Alloys #1-Yb and #3-Hf demonstrate a lower rate of releasing secondary particles (smaller change in Δρ).

When UFG alloys are annealed for 30 minutes, grains begin to grow at 375 °C. No abnormal grain growth was observed (Fig. 4). Analysis of the research findings suggests that the lowest grain growth rate is observed in Alloy #3-Hf, with the average grain size after annealing at 500 °C not exceeding 2.6–3 μm (Fig. 3b). The average grain size in annealed Alloys #1-Yb (Fig. 3a) and #2-Er is close to 4.5–5 μm.

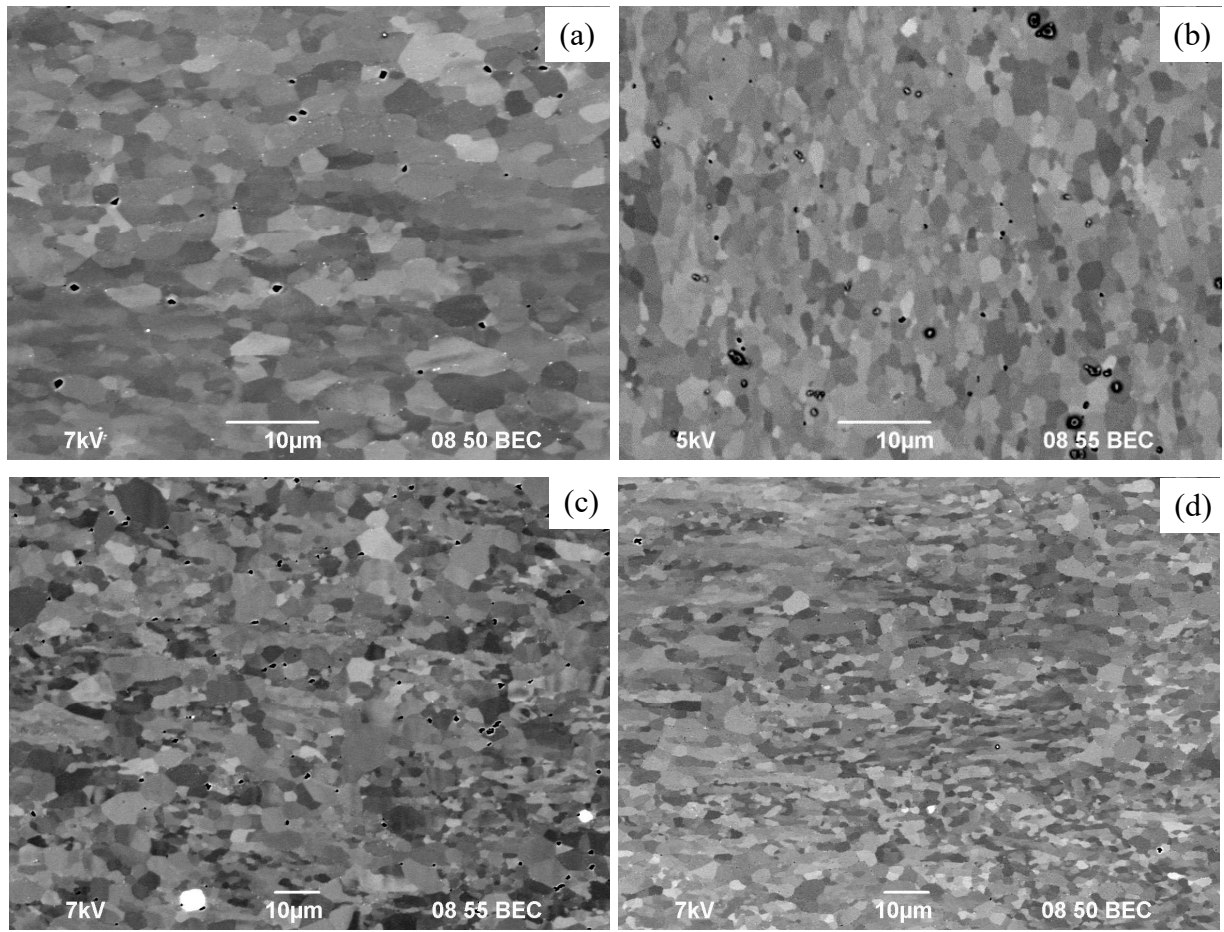

Figure 4. UFG alloys microstructure after annealing at 500 °C, 30 min: (a) Alloy #1-Yb, (b) Alloy #3-Hf, (c) Alloy #4-Zr, (d) Alloy #5-Sc. SEM

Between annealed Alloys #4-Zr and #5-Sc, which are the research objects, the minimum grain size is observed in Alloy #5-Sc. After annealing at 500 ºC (30 min), the average grain size in Alloy #5-Sc is ~2.3 μm. The average grain size in Alloy #4-Zr after annealing at 500 °C is close to 2.8 μm.

The UFG alloys mechanical properties were studied after annealing at the temperature that ensures the maximum hardness of the given alloy (see Fig. 3). The ultimate tensile strength is 340–345 MPa and 385–390 MPa for annealed Alloys #4-Zr and #5-Sc, respectively. Annealed fine-grained Alloys #1-Yb, #2-Er, #3-Hf have an ultimate tensile strength of 375–385 MPa, 360–370 MPa, 345–350 MPa, respectively (Fig. 5). Hence, the mechanical properties of annealed fine-grained alloys are close to Alloy #5-Sc and better than in Alloy #4-Zr (Fig. 5). The fractures of the samples after tensile tests are ductile and represent a set of pits of various sizes (Fig. 6). Strengthening annealing has no significant impact on the nature of fractures. The average pit size in fractures of Alloy #2-Er samples is bigger than in Alloys #1-Yb and #3-Hf samples (Fig. 6). The average pit size in Alloys #1-Yb, #2-Er, #3-Hf is slightly bigger than in Alloy #5-Sc.

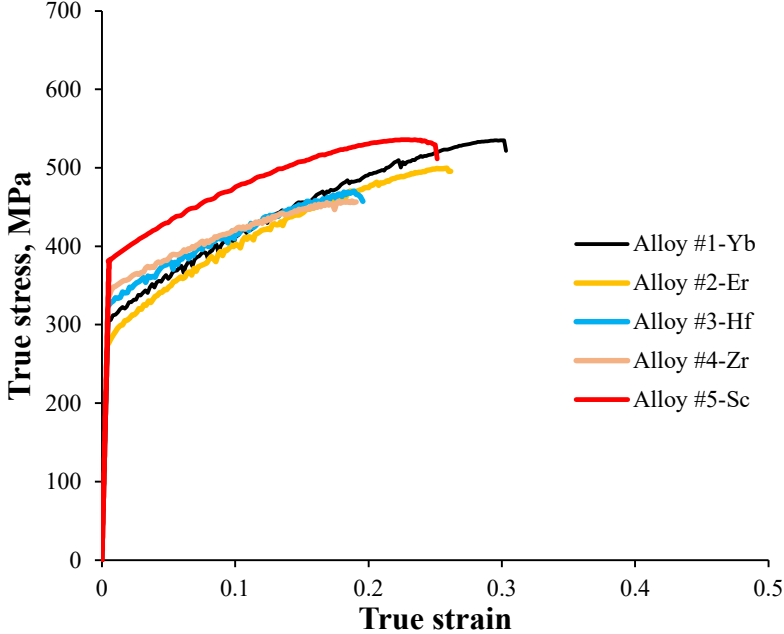

Figure 5. Stress-strain curves σ(ε) for annealed fine-grained alloys with maximum hardness. RT tests at a strain rate of $3.3 \cdot 10^{-3}$ s$^{-1}$

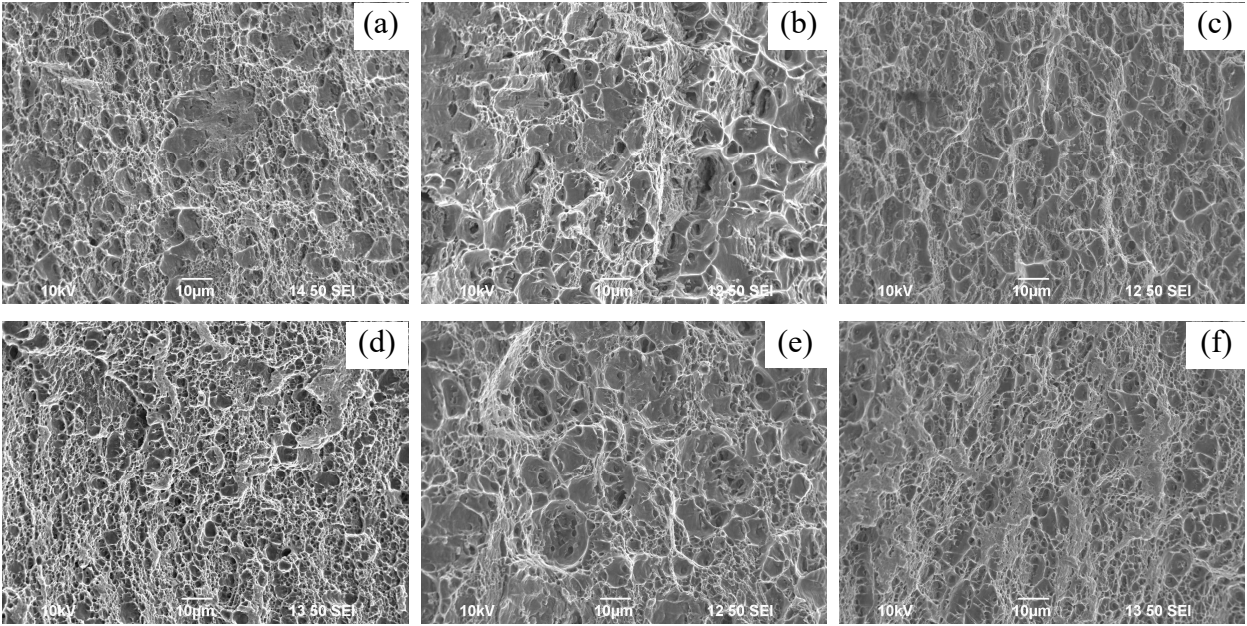

Figure 6. Images of the fracture surface of unannealed UFG alloys (a, b, c) and annealed alloys (d, e, f): Alloys #1-Yb (a, d), #2-Er (b, e), #3-Hf (c, e). Tensile tests at RT. SEM

## 3.2. Deformation behavior at elevated temperatures

Figure 7a shows a typical representation of tensile curves σ(ε) of cast alloy samples at a strain rate of 3.3·10⁻³ s⁻¹. The curves σ(ε) have a typical representation for tension in high-plasticity metals while the curves σ(ε) manifest a stage of stable uniform plastic flow in coarse-grained materials. The uniform strain of cast alloy samples increases as the test temperature goes up.

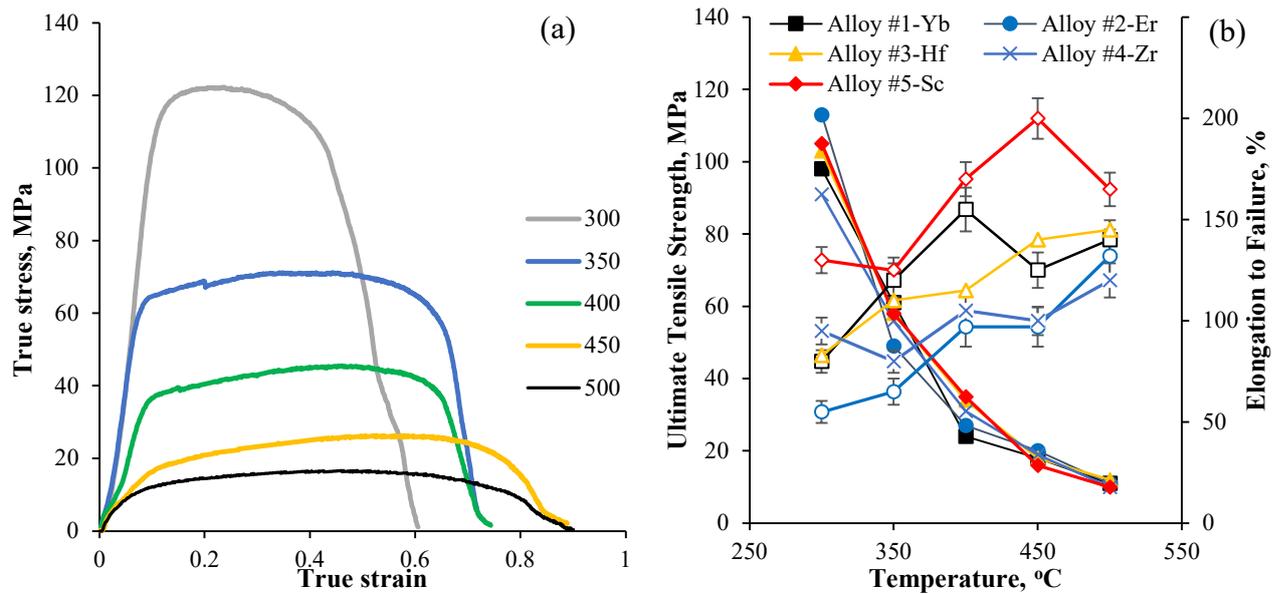

Figure 7. Results of tensile tests of cast alloys at elevated temperatures: (a) curves σ(ε) of cast Alloy #3-Hf; (b) dependence of ultimate tensile strength and elongation to failure on test temperature

As the test temperature increases from 300 to 500 °C, a noticeable decrease in tensile strength from ~90–113 MPa down to 10–12 MPa is observed (Fig. 7b). As the test temperature increases, the ductility of cast alloys goes up, which is well-known. Lower elongation to failure in Alloy #5-Sc at elevated test temperatures (350–450 °C) is driven by precipitation of secondary Al₃(Sc,Zr) particles that hinder the motion of dislocations. The maximum elongation to failure of cast alloys does not exceed 200% (Fig. 7b).

The results of superplasticity tests of UFG alloys are shown in Table 3. Table 3 suggests that plasticity in the UFG alloys is several times higher than in the coarse-grained cast alloys. The highest elongation to failure is observed in UFG Alloy #3-Hf, with strain point at 450 °C and strain rate of

$3.3 \cdot 10^{-2}$ s$^{-1}$, elongation exceeds 1,100% (Table 3). It is important to emphasize that the superplastic properties of UFG Alloy #3-Hf are superior to those of UFG Alloys #4-Zr and #5-Sc with a high content of AEs (in at.%) (Table 1). Alloy #1-Yb manifests good elongation to failure ($\delta$ = 910%) at low temperatures (400 °C).

Table 3. Results of UFG alloys superplasticity tests

| Tensile test mode | | Alloy #1-Yb | | Alloy #2-Er | | Alloy #3-Hf | | Alloy #4-Zr | | Alloy #5-Sc | |
|---|---|---|---|---|---|---|---|---|---|---|---|
| T, °C | $\dot{\varepsilon}$, s$^{-1}$ | $\sigma_b$, MPa | $\delta$, % | $\sigma_b$, MPa | $\delta$, % | $\sigma_b$, MPa | $\delta$, % | $\sigma_b$, MPa | $\delta$, % | $\sigma_b$, MPa | $\delta$, % |
| 300 | $3.3 \cdot 10^{-3}$ | 94 | 260 | 89 | 235 | 55 | 475 | 52 | 670 | 88 | 300 |
| 350 | $3.3 \cdot 10^{-3}$ | 40 | 565 | 58 | 245 | 35 | 815 | 42 | 365 | - | - |
| 400 | $3.3 \cdot 10^{-3}$ | 22 | 910 | 23 | 605 | 31 | 370 | 19 | 690 | 17 | 715 |
| | $3.3 \cdot 10^{-2}$ | 78 | 330 | 48 | 525 | 74 | 350 | 63 | 410 | 35 | 790 |
| | $3.3 \cdot 10^{-1}$ | 137 | 210 | 147 | 210 | 115 | 380 | 99 | 440 | 99 | 415 |
| 450 | $3.3 \cdot 10^{-3}$ | 20 | 255 | 22 | 300 | 20 | 350 | 13 | 830 | 13 | 900 |
| | $3.3 \cdot 10^{-2}$ | 53 | 260 | 47 | 300 | 24 | 1145 | 36 | 345 | 40 | 345 |
| | $3.3 \cdot 10^{-1}$ | 101 | 250 | 80 | 350 | 97 | 270 | 68 | 485 | 80 | 165 |
| 500 | $3.3 \cdot 10^{-3}$ | 14 | 400 | 7 | 565 | 9 | 1000 | 8 | 1005 | 10 | 345 |
| | $3.3 \cdot 10^{-2}$ | 32 | 225 | 25 | 445 | 20 | 300 | 24 | 320 | 26 | 400 |
| | $3.3 \cdot 10^{-1}$ | 52 | 190 | 70 | 250 | 48 | 520 | 46 | 560 | 43 | 865 |

Note: the $\sigma_b$ value was determined using stress-strain curves $\sigma(\varepsilon)$ in the "true strain – true stress" coordinates

Three different patterns of dependence between elongation to failure, strain and strain rate are observed.

Mode I: increase in ultimate tensile strength and decrease in elongation to failure as strain rates go up. This is a common case of strain in highly deformed metals (see [21]). Figure 8 shows a typical representation of stress-strain curves σ(ε) for Mode I of UFG alloys.

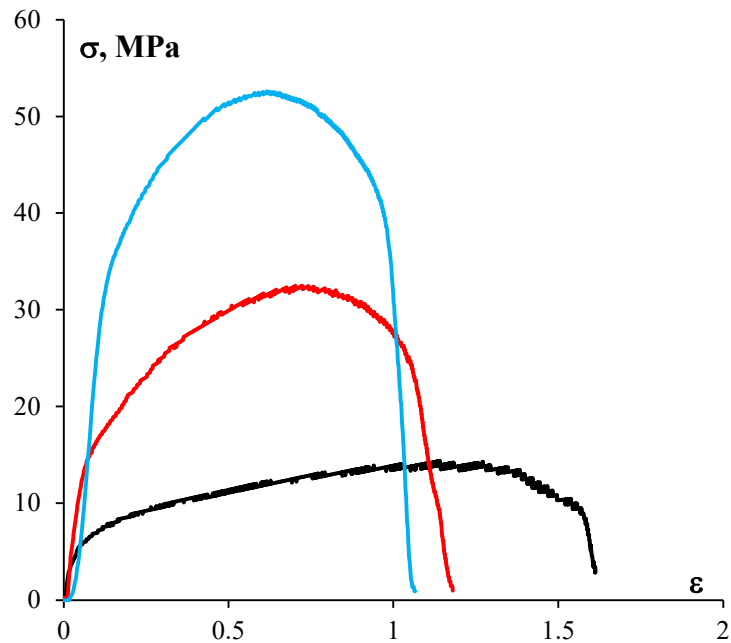

Figure 8. True stress-strain curves σ(ε) of UFG Alloy #1-Yb at 500 ºC. Tests under temperatures and strain rates corresponding to Mode I

Mode II: superplasticity mode, characterized by a non-monotonic dependence between elongation to failure and strain rate; ultimate tensile strength is increasing as strain rates go up. Figure 9 shows a typical representation of stress-strain curves σ(ε) for Mode II deformation of UFG alloys. This pattern of the dependence between elongation to failure and strain rate is classic for superplasticity in fine-grained materials (see [4-6, 18]). Table 2 suggests that Mode II is implemented most often at T ≤ 450 ºC and low strain rates. It should be noted that in some cases the increased ductility stage in UFG alloys is manifested at strain rates under $3.3 \cdot 10^{-3}$ s$^{-1}$. In such cases, an increase in strain rates causes a decrease in ductility of a UFG alloy.

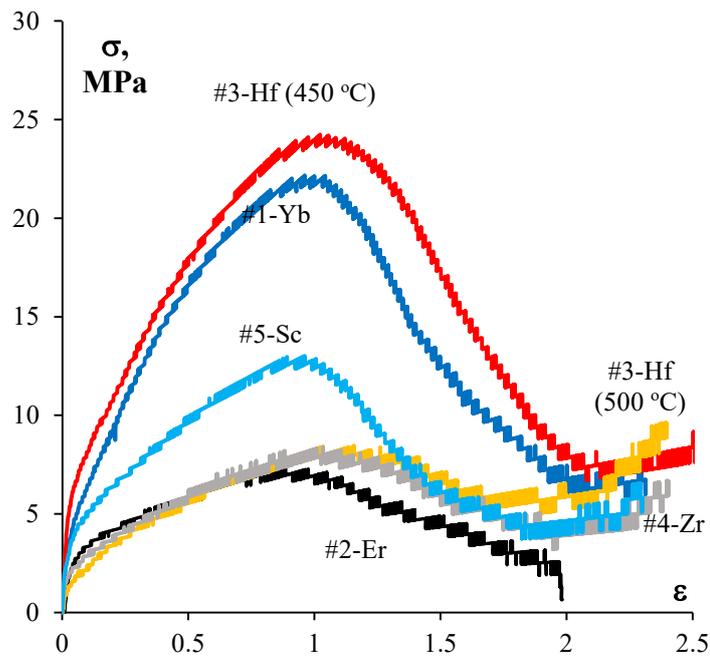

Figure 9. True stress-strain curves σ(ε) of UFG alloys at 500 °C. Tests under temperatures and strain rates corresponding to maximum elongation to failure (Mode II)

Increased yield stress is observed in curves σ(ε) in almost every case where deformation exceeds 800–900%. The only exception is Alloy #2-Er where elongation to failure stays under 600% (Table 3, Fig. 9). This pattern of stress-strain curves σ(ε) is traditionally associated with intense strain-induced (dynamic) grain growth (see [22]). It should be noted that deformation isolation and large necking are not observed until elongation starts. This means that Mode II is mainly uniform plastic deformation of the sample.

It is notable that in some cases there is a simultaneous increase in ultimate tensile strength and elongation to failure as strain rates go up. Fig. 10a shows a typical representation of curves σ(ε) for this deformation of UFG alloys. This effect is most pronounced in UFG Alloys #5-Sc and #3-Hf. In alloys with the addition of Yb and Er, as strain rates go up, a significant increase in ultimate tensile strength is observed with elongation to failure unchanged (Fig. 10b). Table 3 highlights in green the temperature and strain rate conditions under which a simultaneous increase in ultimate tensile strength and ductility is observed.

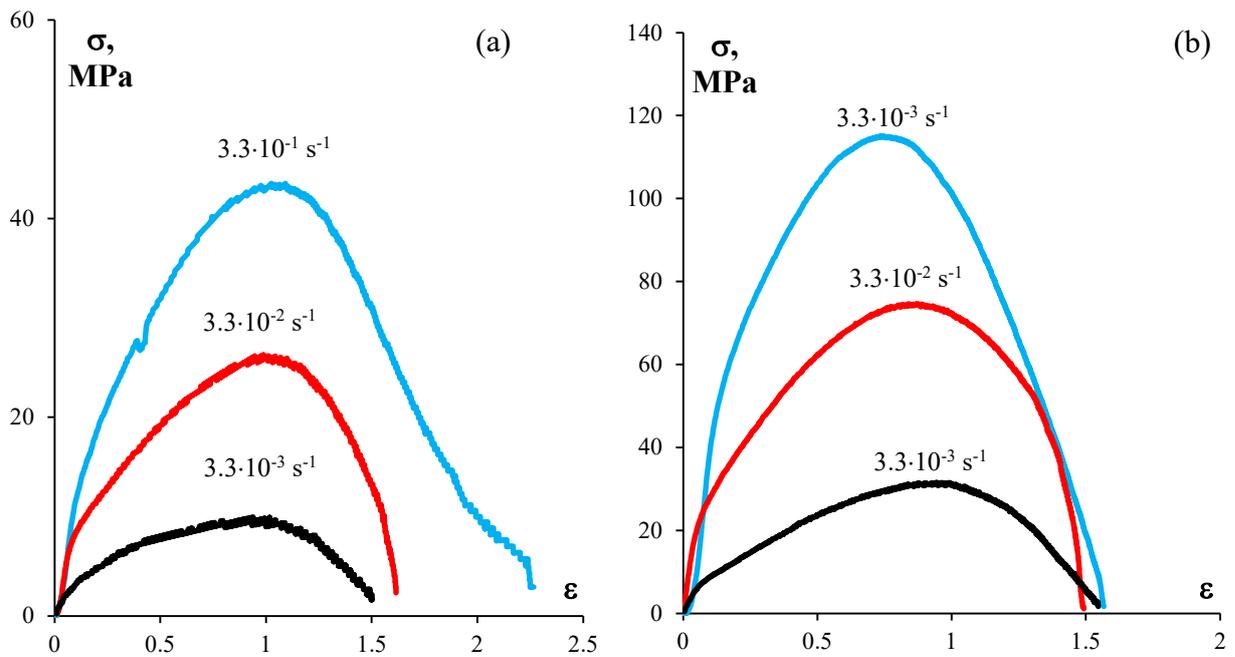

Figure 10. True curves σ(ε) for UFG alloys that are deformed with ductility and ultimate tensile strength increasing simultaneously: (a) Alloy #5-Sc at 500 °C; (b) Alloy #3-Hf at 400 °C

When strain temperatures are high, tensile curves σ(ε) manifest stress peaks indicating dynamic recrystallization that occurs during deformation. When strain temperatures are low, stress peaks in tensile curves σ(ε) are observed mainly at low strain rates. At temperatures of 450 and 500 °C, this type of curves is typical of all strain rates under consideration (Fig. 10b).

Metallographic studies of a samples structure following superplasticity tests suggest that Mode I deformation lends a non-recrystallized microstructure to the alloys. Superplastic deformation in Mode II is accompanied by strain-induced (dynamic) grain growth. As can be seen from Fig. 11, the average grain size in the deformed area is larger than in the undeformed area. This behavior is typical of superplastic materials (see [22-24]).

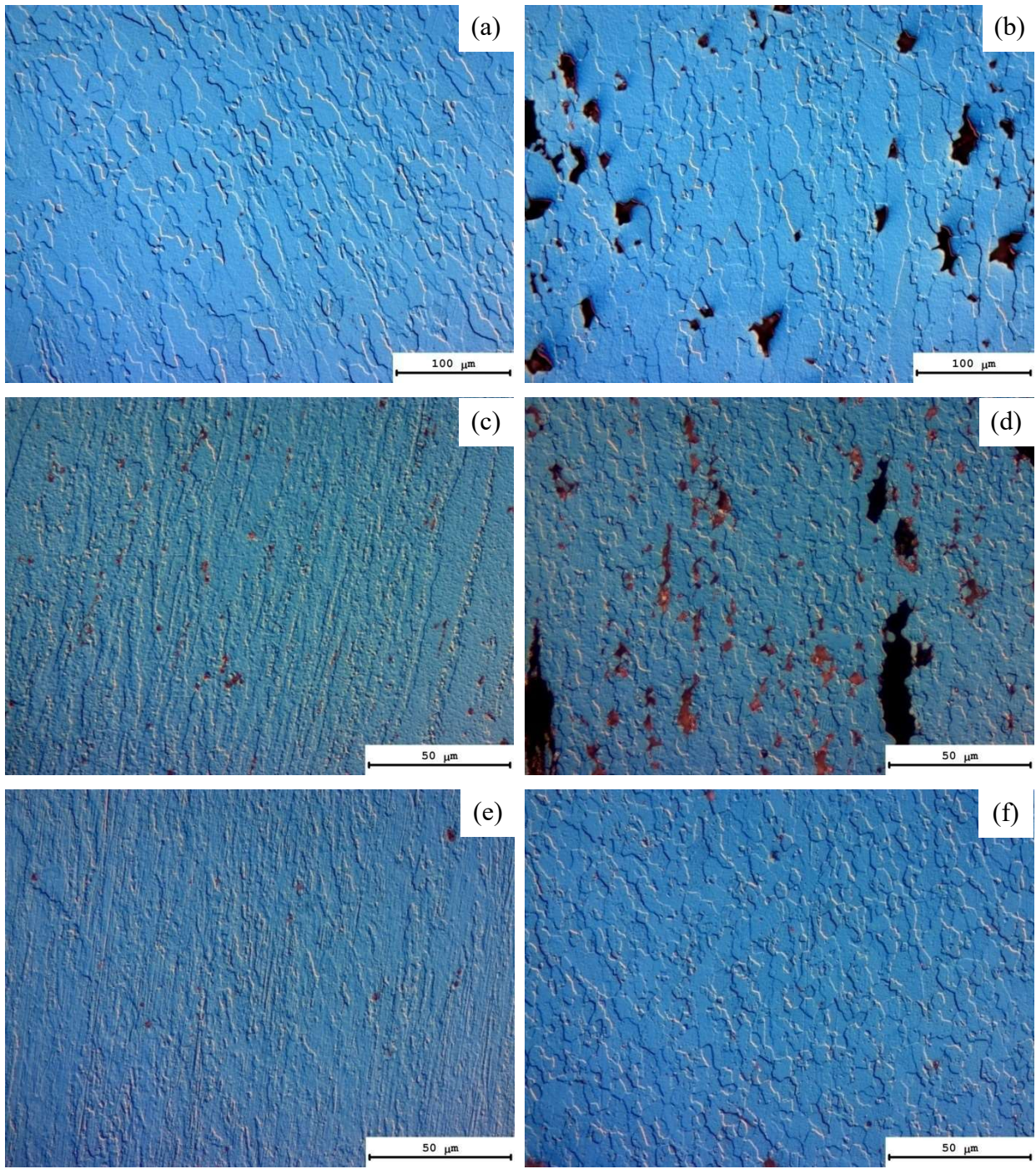

Figure 11. Microstructure of Alloy #1-Yb (a, b), Alloy #2-Er (c, d), Alloy #5-Sc (e, f) after tests at 400 °C (c, d), 450 °C (e, f), 500 °C (a, b). Strain-induced (dynamic) grain growth. Metallography

It should be noted that at 450-500 °C, intense cavitation is observed, causing large pores to appear in the deformed area of the samples. Pores are formed predominantly along the grain boundaries of large recrystallized grains (Fig. 11b, d, f). There are no pores in the undeformed area of the samples. It is also important to emphasize that the size and number of pores in the alloys differ significantly: Alloys #1-Yb and #2-Er formed a significant number of large pores (Fig. 11b, d), Alloys #3-Hf, #4-Zr, #5-Sc manifested single pores (Fig. 11f).

Figure 13 shows the results of studying an alloy samples microstructure after deformation at temperatures and strain rates corresponding to a curve σ(ε) with stress peaks. Fig. 12 shows that the average grain size in the deformed area is smaller than in the undeformed area. The alloys microstructure often has areas where large grains alternate with small ones (Fig. 13). This pattern of microstructural evolution is typical of dynamic recrystallization (see [25-27]).

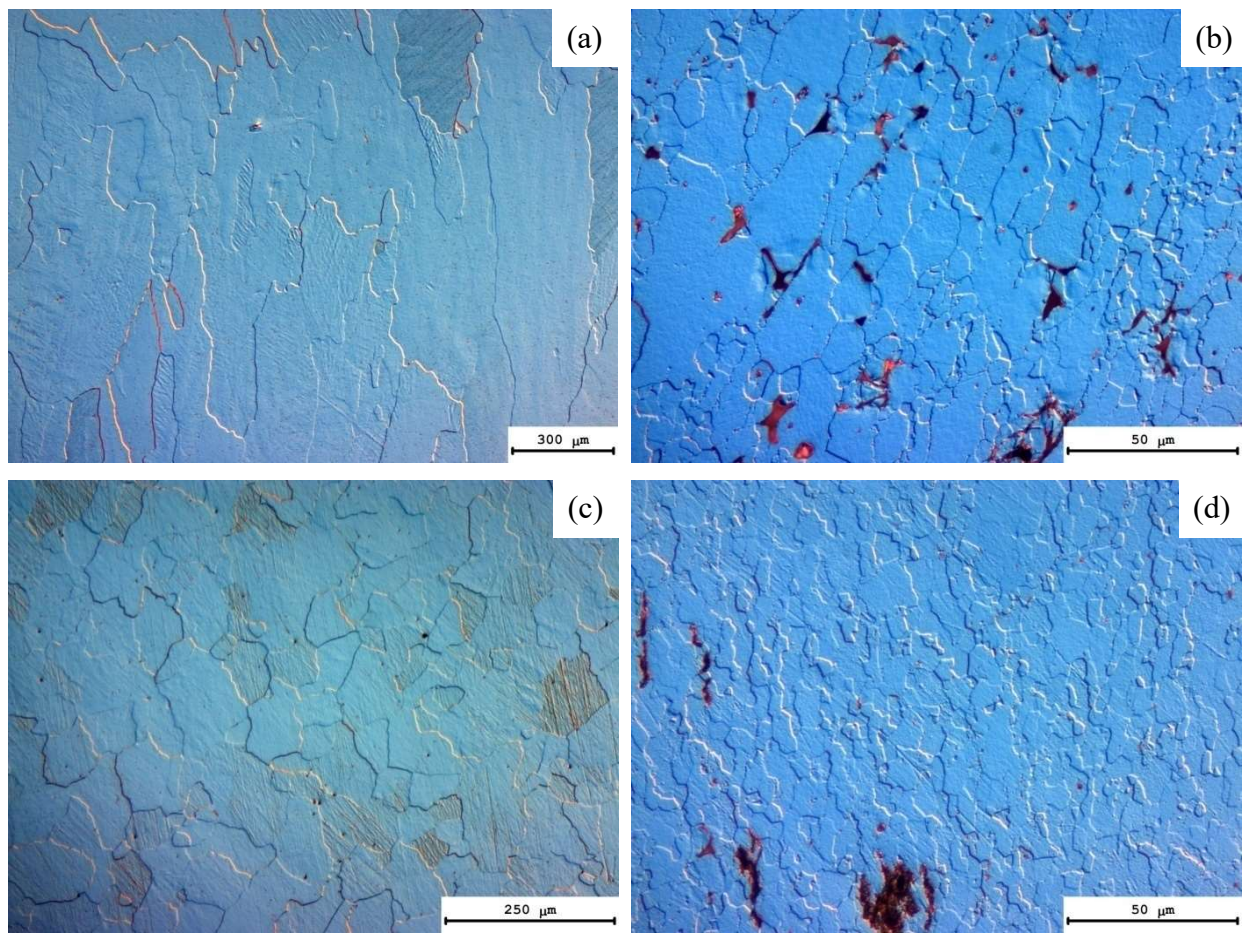

Figure 12. Microstructure of Alloy #1-Yb (a, b) and Alloy #5-Sc (c, d) after superplasticity tests at 450 °C (a, b) and 500 °C (c, d). Dynamic recrystallization. Metallography

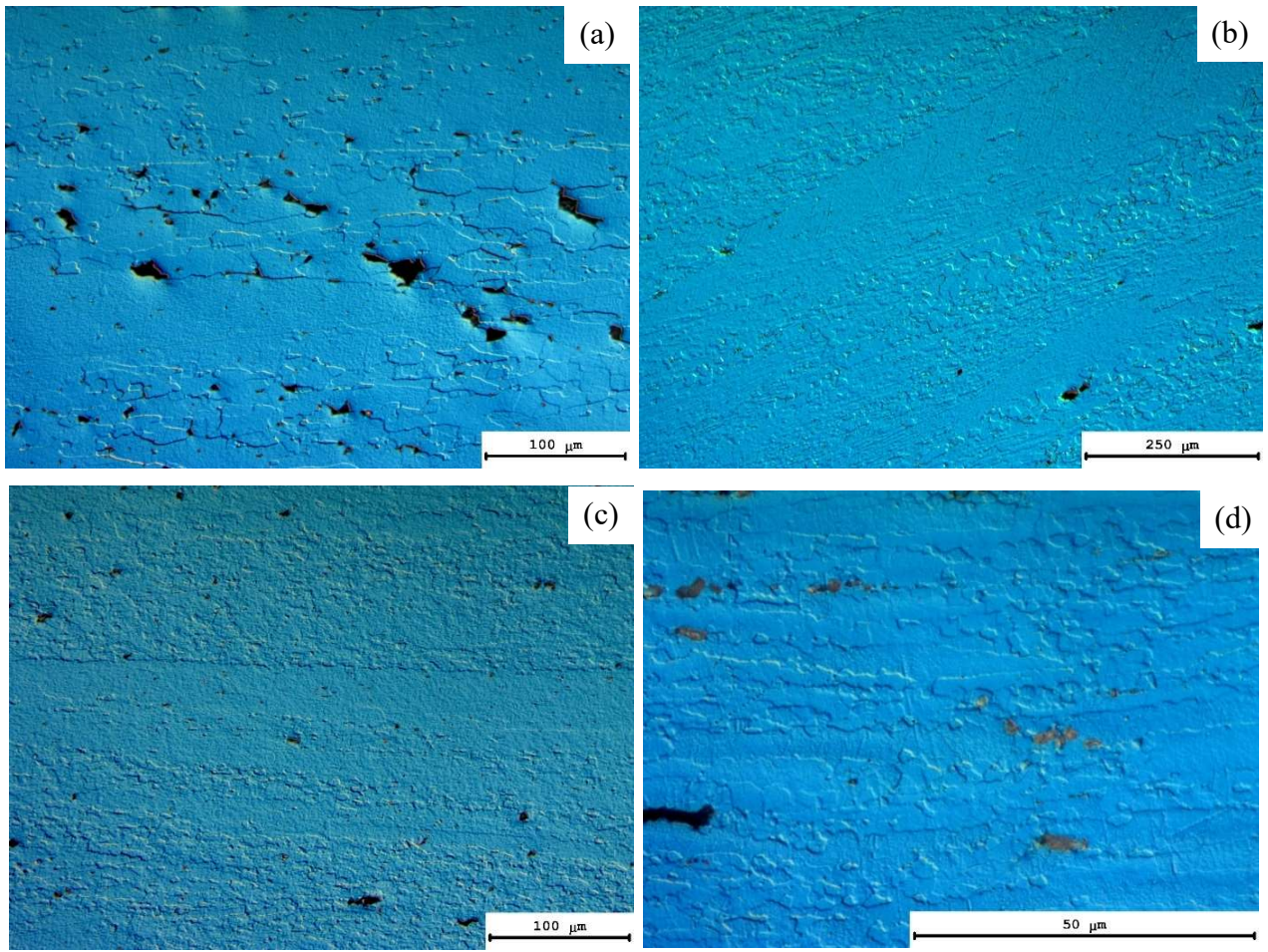

Figure 13. Heterogeneous microstructure of the deformed area of alloys #1-Yb (a), #2-Er (b), #3-Hf (c), #4-Zr (d) after superplasticity tests: (a) 500 °C, $3.3 \cdot 10^{-1}$ s$^{-1}$, (b, d) 500 °C, $3.3 \cdot 10^{-3}$ s$^{-1}$, (c) 450 °C, $3.3 \cdot 10^{-3}$ s$^{-1}$, (d) 400 °C, $3.3 \cdot 10^{-1}$ s$^{-1}$. Metallography

The complex nature of microstructural changes in alloys under superplasticity impacts the type of dependence between hardness and test temperature. Figure 14 shows the results of microhardness measurements in the undeformed area (Fig. 14a) and in the deformed area (Fig. 14b). The dependence between the microhardness of the undeformed area and test temperature has a maximum. The temperature of the dependence maximum Hv(T) is virtually the same as the temperature where the maximum is observed in the dependence between microhardness and the temperature of 30-minute annealing (Fig. 3). This indicates that the first maximum in the dependence between the microhardness of deformed areas and the test temperature is associated with the precipitation of Al$_3$X particles (X = Yb, Er, Hf, Sc, Zr). The second maximum in the Hv(T) dependence corresponds to the strain temperature range where dynamic recrystallization, leading to a smaller average grain size, is observed.

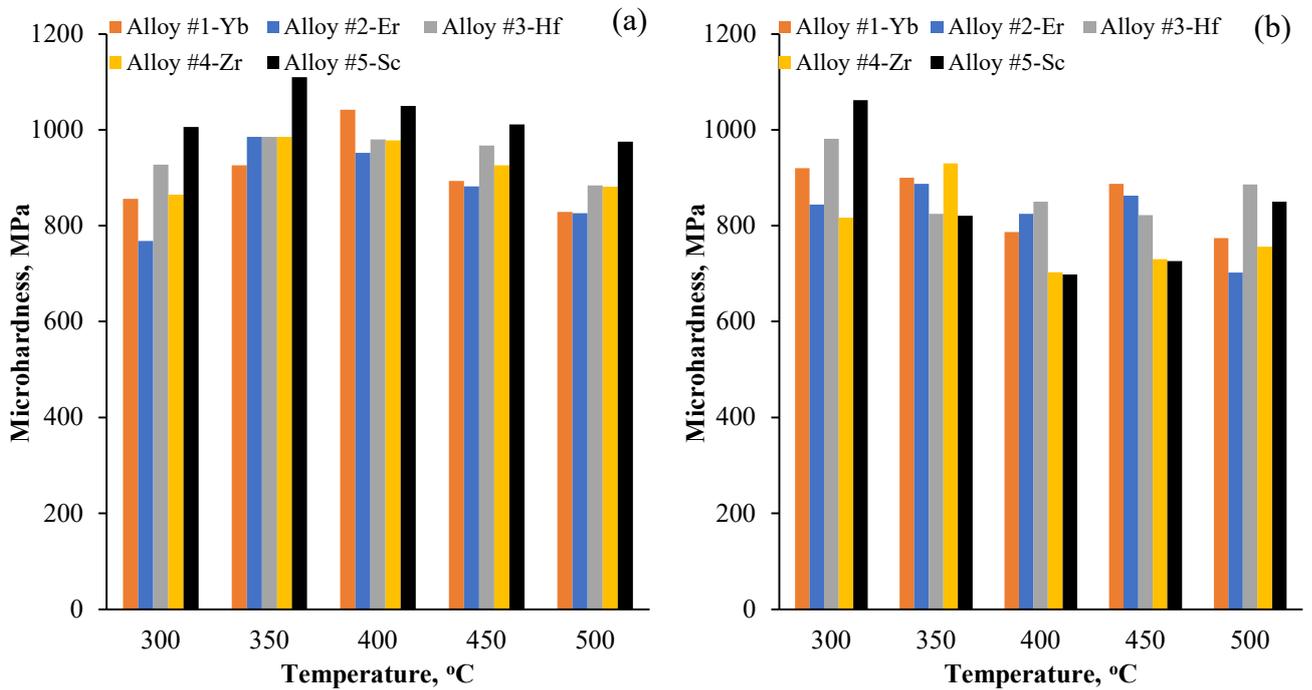

Figure 14. Microhardness (Hv, MPa) in the undeformed area (a) and in the fracture area (b) of aluminum alloy samples after superplasticity tests

SEM demonstrates that fractures in the samples of Alloys #1-Yb, #2-Er, #3-Hf after tensile tests at temperatures of 300-400 °C are typically ductile. The fractures of the samples after testing at 300-400 °C represent a set of pits of various sizes (Figs. 15a, b, c). The central part of the fractures after testing at 450 and 500 °C is a set of pits of various sizes, alternating with conchoidal fracture areas (Figs. 15d, e, f). This type of fracture is common for ruptures occurring simultaneously along grain boundaries and sweeping planes. Raising the test temperature from 450 to 500 °C leads to an increase in the surface of a conchoidal fracture area and bigger size of the generic fracture elements.

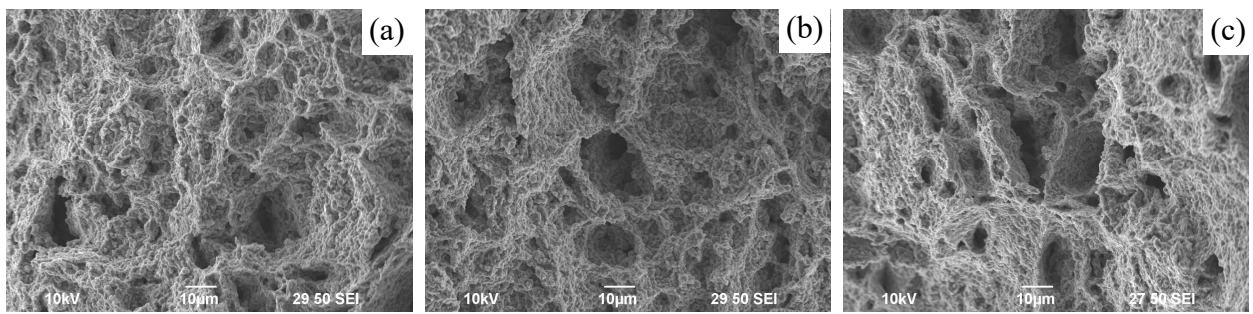

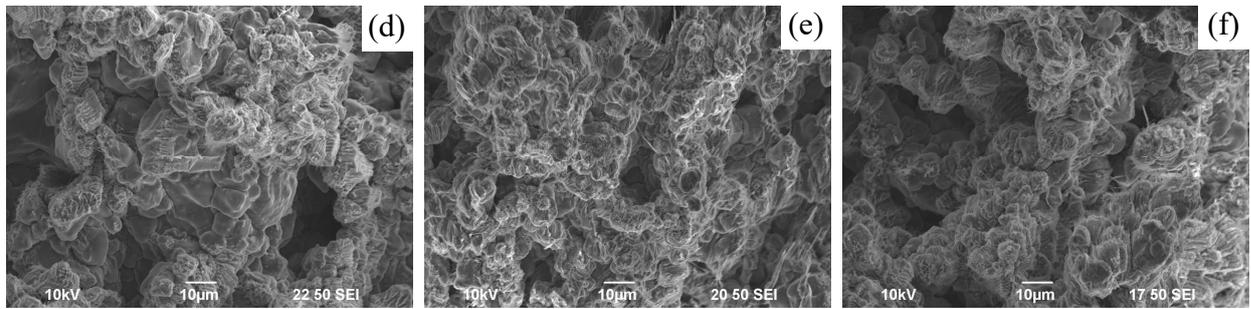

Figure 15. Fracture surface of Alloys #1-Yb (a, d), #2-Er (b, e), #3-Hf (c, f) after tensile tests at 300 °C (a, b, c) and 500 °C (d, e, f). Strain rate of $3.3 \cdot 10^{-3}$ s$^{-1}$

A comparison of the results of fractographic analysis of fractures with the results of tensile tests (Table 3, Figs. 8, 9, 10) indicates that different deformation modes are characterized by different fracture patterns. Fractures of alloy samples deformed in Mode II are characterized by pits of various sizes while fractures of samples deformed in the dynamic recrystallization mode manifest a set of pits and conchoidal fracture areas.

3.3 Analysis of stress-strain curves σ(ε)

The value of strain hardening exponent $n = \partial \ln(\sigma)/\partial \ln(\varepsilon)$ is an important characteristic that describes the tendency of a material to undergo uniform deformation. As per the model [28], the value of the exponent $n$ depends on the mechanism of defect accumulation at the grain boundaries under superplastic deformation.

Figure 16 demonstrates curves σ(ε) in logarithmic coordinates ln(σ)–ln(ε), the slope of which corresponds to the value of the strain hardening exponent *n*. It can be observed from Fig. 16a that the curves σ(ε) in logarithmic coordinates are three-staged, the value of the exponent *n* depending on the strain and test temperatures. At 300 °C and low strain degrees, high values of the exponent *n* ~ 1 are observed; for higher strain degrees that correspond to the stage of stable uniform flow, the value of *n* is close to *n* ~ 0.5. When the deformation temperature rises to 450–500 °C, the first stage of intense strain hardening with a large exponent *n* is insignificant. Increasing strain rates raise the exponent *n* during the first stage of strain hardening but have no significant impact on the intensity of strain hardening during the second stage of the curve σ(ε) (Fig. 16b). The value of the strain hardening exponent at elevated strain rates is close to $n \sim 0.3$. Alloys #1-Yb, #2-Er, #3-Hf achieve the maximum values of the exponent $n$ at test temperatures of 400-450 °C while the values of the exponent $n$ decrease at temperatures of 300-350 and 500 °C (Fig. 17a). The highest value of the strain hardening exponent ($n \sim 0.60$-$0.61$) is observed under superplastic deformation of UFG Alloy #3-Er at 400 °C (Fig. 17a). Alloys #4-Zr and #5-Sc achieve the highest values of the exponent $n$ at elevated temperatures and strain rates (Fig. 17b).

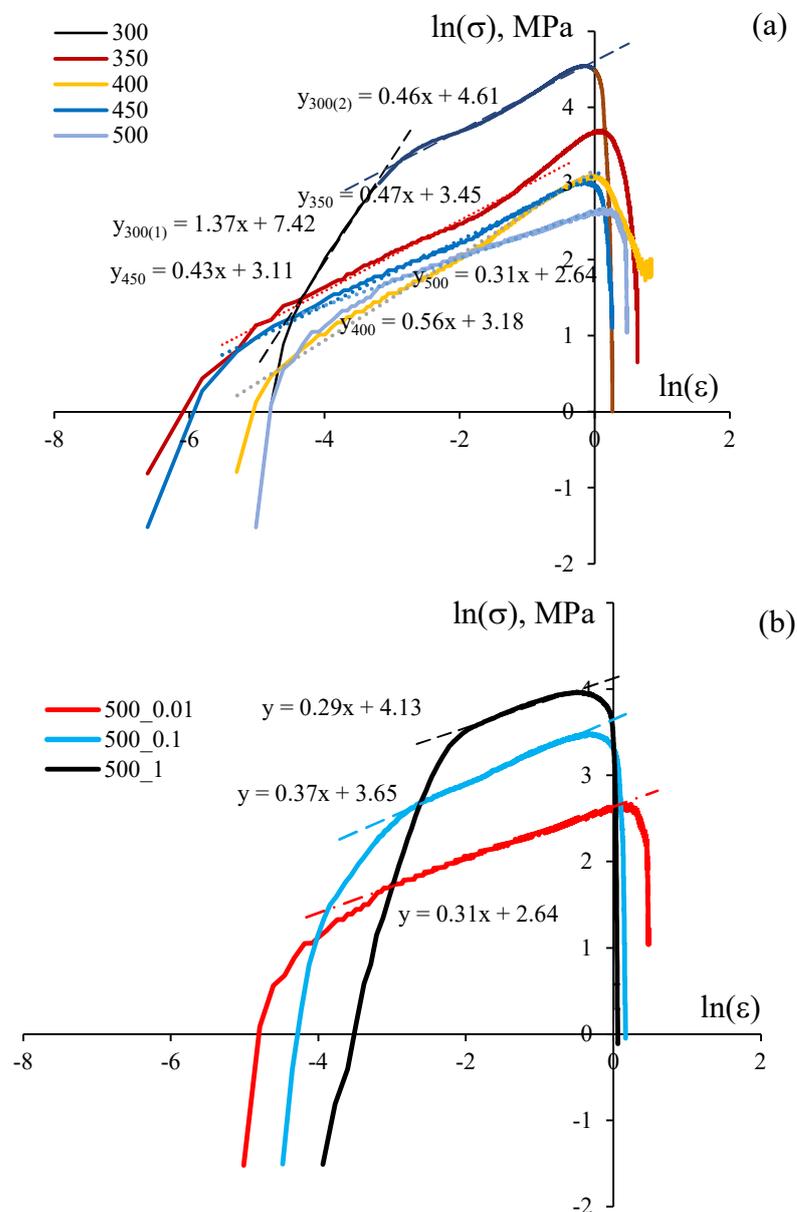

Figure 16. Curve σ(ε) in logarithmic coordinates for UFG Alloy No.1-Yb: (a) tests at various temperatures, at a strain rate of $3.3 \cdot 10^{-3}$ s$^{-1}$; (b) tests at various strain rates, at 500 °C

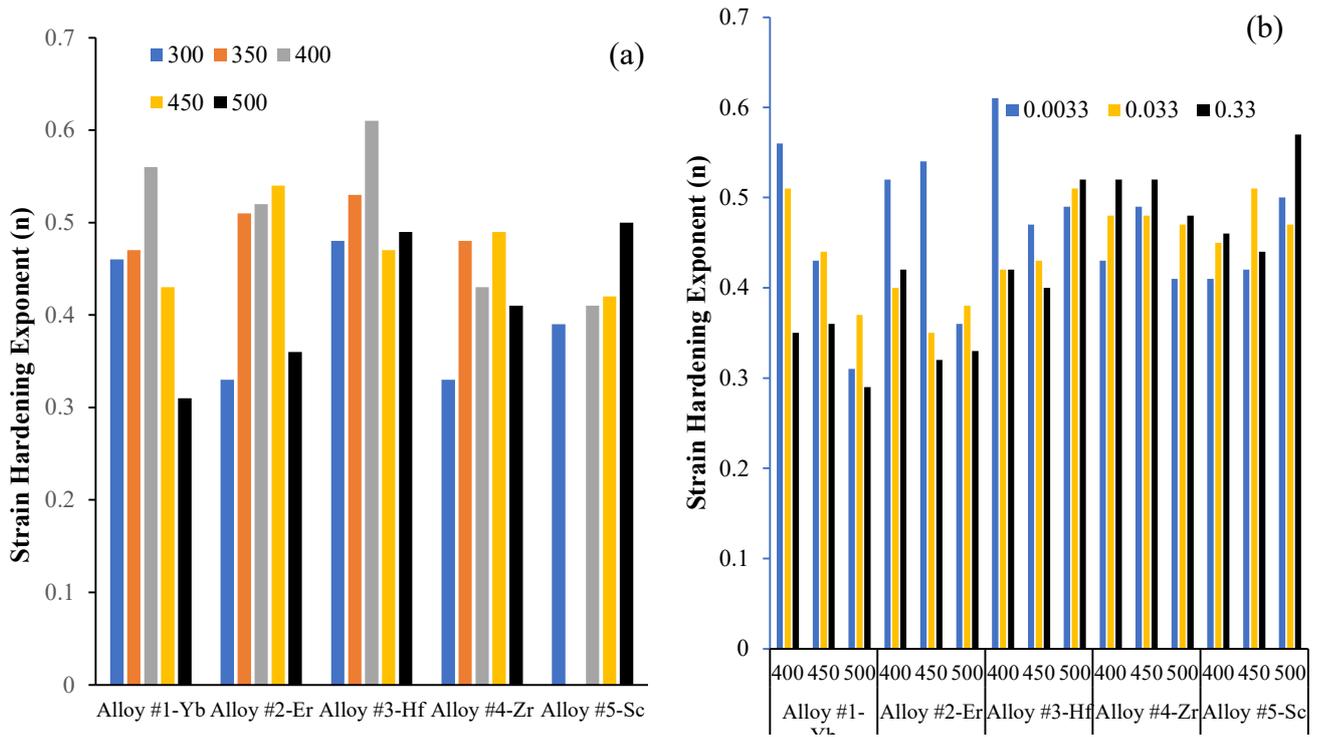

Figure 17. Values of the strain hardening exponent *n* for UFG alloys at different temperatures (a) and strain rates (b)

The value of threshold flow stress ($\sigma_P$) is an important property driven by microstructural parameters of a UFG alloy. The techniques for determining the $\sigma_p$ value are described in detail in [29, 30]. As per [30], the value $\sigma_p$ can be determined using the dependences $\sigma_b - \dot{\varepsilon}^{1/k}$, where *k* is the numerical factor, the physical concept of which corresponds to the rate sensitivity factor of yield stress $k = 1/m$ (see Eq. (1)). The point of intersection with the stress axis at zero strain rate ($\dot{\varepsilon}^{1/k} = 0$) correponds to the threshold stress value $\sigma_p$. The value of the factor *k* for the alloys under consideration was determined using the least squares technique for the slope of the $\ln(\sigma_b) - \ln(\dot{\varepsilon})$ dependence. Figure 19a features the dependences $\sigma_b - \dot{\varepsilon}^{1/k}$ for the UFG alloys being studied.

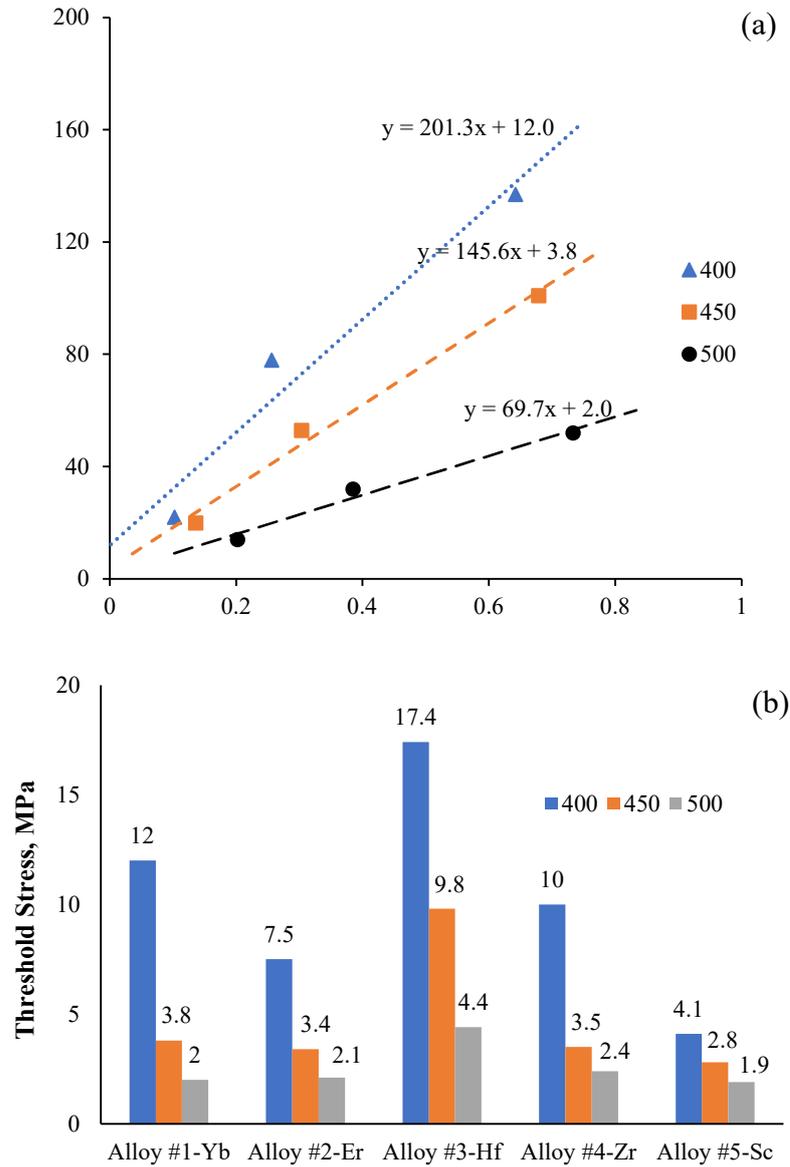

Figure 18. Dependences between stress and strain rates in the coordinates $\sigma_b - \dot{\varepsilon}^{1/k}$ for UFG Alloy #1-Yb (a) and threshold stress value for UFG alloys at different test temperatures (b)

Fig. 18a suggests that raising the test temperature leads to a lower threshold stress $\sigma_p$. This result fits well with the data of [30]. The analysis of the data in Fig. 18b indicates that the highest threshold stress is observed in Alloy #3-Hf while Alloy #5-Sc manifests the lowest threshold stress.

As previously detailed, the value of the rate sensitivity factor of yield stress (*m*) can be determined through the slope of the dependence $\ln(\sigma_b) - \ln(\dot{\varepsilon})$ (see Eq. (1)). As per [18], the optimal superplastic deformation mode has a value of *m* = 0.5. Analysis of the data in Fig. 19 indicates that the value of the factor *m* for Mode II superplastic deformation is *m* ~ 0.3–0.4.

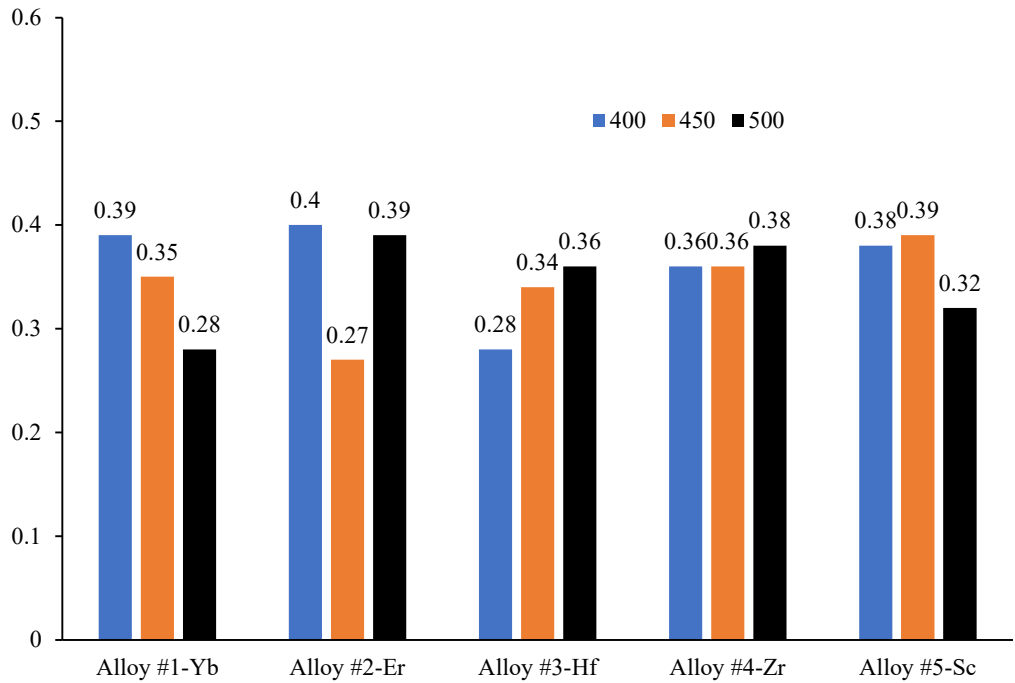

Figure 19. Values of the factor *m* for UFG alloys at different temperatures

## 4. Discussion

4.1. Analysis of superplastic deformation mechanisms

First, the general patterns of deformation behavior of UFG Al-Mg-Sc-Zr-(Yb, Hf, Er) alloys will be described.

The research results demonstrate that there is a competition between the processes of strain-induced grain growth and dynamic recrystallization during superplastic deformation of UFG alloys (see also [31]). At low heating temperatures, strain-induced grain growth predominantly occurs while, at elevated temperatures, dynamic recrystallization is observed. Dynamic recrystallization is followed by grain refinement in the deformed area of the sample and a change in the nature of the curve $\sigma(\varepsilon)$. It should be noted that grains retain a near-equiaxial shape during superplastic deformation (Figs. 11, 12). It prompts a suggestion that grain boundary sliding (GBS) is the key mechanism of superplastic deformation. This assumption is supported by very high values of the strain rate sensitivity factor (*m* ~ 0.5).

The results of microstructural studies of alloys following superplasticity tests show that dynamic recrystallization begins with an increase in the average grain size to a critical value of $d^* \sim$ 5–10 μm. When the average grain size goes below the critical size of $d < d^*$ the contribution of dynamic recrystallization is reduced and the contribution of strain-induced grain growth increases. The rate of grain boundaries migration under strain-induced grain growth depends on temperatures and strain rates [22]. Minor elongations to failure (short test time) at 400 °C and $\dot{\varepsilon} \geq 3.3 \cdot 10^{-2}$ s$^{-1}$ cause an insignificant grain growth ($d < d^*$). This is likely to exclude dynamic recrystallization at 400 °C and strain rates of $\dot{\varepsilon} \geq 3.3 \cdot 10^{-2}$ s$^{-1}$.

In order to mitigate the causes of softening under superplasticity of UFG alloys, one can resort to an approach based on the concept whereby stable uniform plastic flow stops when the speed or strain hardening, characterized by factors $m$ and $n$, respectively, becomes insufficient. As per Hart's criterion, the condition for stopping stable plastic flow of a material under tension can be represented as [32]:

$$m + n/\varepsilon \geq 1. \qquad (2)$$

(2) implies that the plastic flow of the material can remain uniform up to strain $\varepsilon_1$:

$$\varepsilon_1 \leq n / (1 - m). \qquad (3)$$

Experimental data of factors $n$ (Fig. 17) and $m$ (Fig. 19) will be used to determine deformation $\varepsilon_1$ where localization of plastic flow begins, which is reflected in the curve $\sigma(\varepsilon)$ as a softening stage. Table 4 presents calculation results and their comparison with experimental data. The average error of measuring the experimental value $\varepsilon^*$ is 0.02-0.03.

Table 4. Experimental and theoretical values of uniform deformation at a $3.3 \cdot 10^{-3}$ s$^{-1}$

| T, °C | $\dot{\varepsilon}$, s$^{-1}$ | Alloy #1-Yb | | Alloy #2-Er | | Alloy #3-Hf | | Alloy #4-Zr | | Alloy #5-Sc | |
|---|---|---|---|---|---|---|---|---|---|---|---|
| | | $\varepsilon^*$ | $\varepsilon_1$ | $\varepsilon^*$ | $\varepsilon_1$ | $\varepsilon^*$ | $\varepsilon_1$ | $\varepsilon^*$ | $\varepsilon_1$ | $\varepsilon^*$ | $\varepsilon_1$ |
| 400 | $3.3 \cdot 10^{-3}$ | 1.05 | 0.92 | 0.94 | 0.87 | 1.07 | 0.85 | 1.07 | 0.67 | 0.95 | 0.66 |
| | $3.3 \cdot 10^{-2}$ | 0.81 | 0.84 | 1.09 | 0.67 | 0.88 | 0.58 | 0.93 | 0.80 | 1.04 | 0.73 |
| | $3.3 \cdot 10^{-1}$ | 0.59 | 0.57 | 0.65 | 0.70 | 0.77 | 0.58 | 0.86 | 0.81 | 0.88 | 0.74 |
| 450 | $3.3 \cdot 10^{-3}$ | 0.88 | 0.66 | 0.96 | 0.74 | 1.00 | 0.71 | 1.04 | 0.77 | 1.00 | 0.69 |
| | $3.3 \cdot 10^{-2}$ | 0.81 | 0.68 | 0.81 | 0.48 | 1.10 | 0.65 | 1.00 | 0.75 | 0.98 | 0.84 |
| | $3.3 \cdot 10^{-1}$ | 0.70 | 0.55 | 0.76 | 0.44 | 0.74 | 0.61 | 0.89 | 0.81 | 0.76 | 0.72 |
| 500 | $3.3 \cdot 10^{-3}$ | 1.29 | 0.43 | 1.00 | 0.59 | 1.32 | 0.77 | 1.04 | 0.66 | 1.05 | 0.74 |
| | $3.3 \cdot 10^{-2}$ | 0.76 | 0.51 | 0.93 | 0.62 | 0.94 | 0.80 | 0.94 | 0.77 | 1.01 | 0.69 |
| | $3.3 \cdot 10^{-1}$ | 0.64 | 0.40 | 0.69 | 0.54 | 0.99 | 0.81 | 0.98 | 0.84 | 1.01 | 0.84 |

Note: $\varepsilon^*$ stands for the terminal value of uniform deformation determined through analysis of the tensile curves $\sigma(\varepsilon)$ in the "true stress – true strain" coordinates.

It can be observed from Table 4 that the calculated values of the uniform strain $\varepsilon_1$ are lower than the experimental value $\varepsilon^*$. The best match between the values $\varepsilon_1$ and $\varepsilon^*$ can be achieved at high strain rates. The divergence of the experimental and theoretical values of uniform deformation can be explained, first of all, by the need to use the value $m$ averaged over all strain rates when making calculations using Eq. (2). The second reason is a very smooth nature of the curves $\sigma(\varepsilon)$ at low strain rates, which makes it difficult to define the value $\varepsilon^*$ well. It is also important to emphasize that, unlike magnesium-free UFG alloys [23, 24], the UFG Al-6%Mg-Sc-Zr-(Yb,Er,Hf) alloys being studied do not manifest a pronounced susceptibility to necking and localization of deformation.

The nature of UFG alloys destruction will be analyzed below.

The results of microstructural studies demonstrate that pores form and grow rigorously during superplasticity tests. Pores are formed most intensely when testing samples of UFG Alloys #1-Yb and #2-Er for superplasticity.

The model of cavitation destruction of fine-grained alloys under superplasticity is described in [33]. As per [33], under superplasticity, pores are formed in secondary particles at triple junctions of grains. It is difficult for dislocations to cut through such particles and disclination defects are formed around them under superplastic deformation. As per [33], a defect formed during deformation in a particle of R radius located in the grain boundary can be described to a first-order approximation as a disclination loop of radius R and power ω(τ). The power of the disclination loop ω(τ) grows proportionately to the number of defects occurring at the boundary. When defects occur at grain boundaries due to intragranular deformation going at the rate of $\dot{\varepsilon}_v$, the power of the disclination loop at the initial stage of its formation can be calculated using the formula $\omega(t) = \psi_1 \dot{\varepsilon}_v \tau$ [33], where $\psi_1$ is the geometric factor and $\dot{\varepsilon}_v$ is the rate of intragranular deformation. As defects accumulate at grain boundaries, the power of the disclination loop ω(t) increases, and so does the elastic energy associated with this defect. At a certain value of critical power ω* the excess energy of the loop becomes so high that it becomes energetically advantageous for the grain boundary to "free itself" from the source of this energy. At elevated deformation temperatures, such relaxation of the energy stored can occur through formation of micropores at the particle-aluminum boundaries. During superplastic deformation, pores will grow in proportion to the degree and strain rate [22] and once reaching their critical size, such pores become a source of UFG alloy fracture.

Large primary Al$_3$X particles formed during crystallization of alloys under superplastic deformation can cause micropores to form and, consequently, be the cause of cavitation fracture of UFG alloys. Hence, in our opinion, the reason for different values of the maximum elongation to failure of UFG alloys of different compositions is, first of all, large primary particles in the material structure. As is shown in Section 3.1, micron-sized primary particles are observed in alloys containing Er and Yb, uniformly distributed throughout the volume of the material (Fig. 1d). Large primary particles were not detected in Alloys #3-Hf, #4-Zr, #5-Sc, which ensures their better ductility at high test temperatures (Table 3).

It should be noted that, secondary particles can precipitate during heating of UFG aluminum alloys, growing quite quickly and also causing micropores to form. In our opinion, precipitation of secondary $Al_3X$ nanoparticles can cause single micropores to form in samples of UFG Alloys #3-Hf, #4-Zr, #5-Sc after superplasticity tests (Fig. 11e, 12d). It should be noted that secondary nanoparticles are much smaller than primary nanoparticles, therefore the contribution of secondary nanoparticles precipitating during heating to the cavitation destruction of UFG alloys is considerably less.

In conclusion, the reasons for a simultaneous increase in flow stress and ductility of UFG alloys should be discussed.

As per [28, 34], during superplastic deformation of UFG alloys, defects are accumulated at grain boundaries: orientation misalignment dislocations and sliding (tangential) components of the Burgers vector of delocalized dislocations. The magnitude of the flow of defects entering the grain boundaries is proportional to the strain rate [28, 34]. As defects are accumulated at grain boundaries, a long-range internal stresses from grain boundaries is formed, its magnitude being proportional to the defect density [28, 34]. In addition, dislocations occurring in grain boundaries increase their free (excess) volume and grain-boundary diffusivity [28]. This allows for higher rates of GBS during superplasticity and of fragmentation and grain boundaries migration under dynamic recrystallization. This leads to a higher yield stress and, concurrently, more intense uniform deformation of the fine-grained material.

4.2. Analyzing the alloying impact on UFG alloys superplastic behavior

The effect of AEs (Yb, Hf, Er) on superplastic behavior of UFG aluminum Al-6Mg-Sc-Zr alloy will be analyzed below. Before analyzing the results, two important aspects should be noted.

First, it is necessary to take into account the nature of the spatial distribution of AEs during melt crystallization, which can be described using the distribution coefficient $K = C_S/C_L$, where $C_S$ and $C_L$ represent concentrations of the AEs in the solid phase and in the melt, respectively. With K > 1, AEs are concentrated predominantly throughout the grains volume (volume of growing dendritic crystals), and with K < 1, they are concentrated at the grains/dendritic boundaries. For example, the distribution coefficient of scandium in aluminum is $K_{Sc(Al)}$ = 0.9–1 [35] and, therefore, Sc is uniformly distributed throughout the crystallized metal volume. The distribution coefficient of Zr in Al is $K_{Zr(Al)}$ = 2.5–2.54 [35] and, therefore, Zr is concentrated mainly in the grains volume during crystallization. The distribution coefficient of Yb in Al is very small $K_{Yb(Al)}$ = 0.08 [35].

Second, it should be noted that the AEs (Yb, Er, Hf vs. Sc, Zr) have different atomic weights: 173 g/mol for Yb, 167 g/mol for Er, 178 g/mol for Hf, 45 g/mol for Sc, 91 g/mol for Zr. Therefore, replacing 0.1 wt.% Zr or 0.1 wt.% Sc with 0.1 wt.% Yb, Er, Hf decreases the total content of AEs in at.% (Table 1 ).

The maximum content of AEs in Alloy #5-Sc (Table 1) determines a large volume fraction of precipitated $Al_3$(Sc,Zr) particles and, consequently, high strength and hardness of Alloy #5-Sc after annealing (Figs. 3, 5). The maximum elongation to failure of Alloy #5-Sc is not very high (865-900%) (Table 3). It is notable that Alloy #4-Zr has a lower content of AEs in at.% (Table 1) but demonstrates higher superplastic characteristics, with elongation to failure reaching 1,000% at 500 °C (Table 3).

The concentration of Sc is much higher than the concentration of Zr in Alloy #5-Sc (Sc:Zr = 4.3 at.%), and it can be assumed that, when heating Alloy #5-Sc, predominantly $Al_3$Sc particles in the volume of grains precipitate while there is a small number of $Al_3$(Sc$_{0.75}$Zr$_{0.25}$) particles along the grain boundaries of the aluminum alloy.

In Alloy #4-Zr, the total concentration of Zr (in at.%) is close to the concentration of Sc in at.% (Sc:Zr = 1.2), but the local concentrations of Sc and Zr throughout the volume of grains and at grain boundaries can vary significantly. Taking $K_{Sc(Al)}$ = 1 [35] with the total concentration of Sc being 0.072 at.%, it can be assumed that ~ 0.036 at.%Sc is concentrated in the grain boundaries and the same amount of Sc is concentrated throughout the volume of the grains. The grain boundaries of Alloy #4-Zr contain an increased concentration of Zr (~0.043 at.% with $K_{Zr(Al)}$ = 2.5) and ~0.036

at.%Sc (see above). Since the concentrations of Sc and Zr in the grain boundaries of Alloy #4-Zr are close, it can be assumed that $Al_3(Sc_{0.5}Zr_{0.5})$ particles predominantly precipitate in the grain boundaries of Alloy #4-Zr. $Al_3(Sc_{0.75}Zr_{0.25})$ particles precipitate throughout the volume of Alloy #4-Zr (~0.036 at.% Sc, ~0.017 at.% Zr) grains.

It has been demonstrated in [36, 37] that $Al_3(Sc_xZr_{1-x})$ particles have an "$Al_3Sc$ core – $Al_3Zr$ shell" structure. This allows for an efficient combination of rapidly releasing $Al_3Sc$ particles at low temperatures and high thermal stability of the $Al_3Zr$ phase at elevated heating temperatures. This makes it possible to ensure a high thermal stability of Alloy #4-Zr microstructure despite the content of AEs in it being lower than in Alloy #5-Sc. As demonstrated above, the average grain size in Alloys #4-Zr and #5-Sc after annealing at 500 °C is ~2.8 μm and ~2.3 μm, respectively.

A high thermal stability of $Al_3(Sc_{0.5}Zr_{0.5})$ particles helps to maintain a small grain size in Alloy #4-Zr under superplastic deformation and to ensure high elongation to failure. This result correlates well with higher values of the threshold stress in Alloy #4-Zr compared to Alloy #5-Sc (Fig. 18), although these alloys have similar values of the strain hardening exponent $n$ (Figs. 16, 17) and similar values of the rate sensitivity factor $m$ (Fig. 19). Hence, partial replacement of scandium with zirconium is rather an efficient way to ensure high superplastic characteristics of UFG Al-Mg alloys. This result is in good agreement with the available literature data [5, 6] and the case history of using industrial alloys of Al-15XX series (alloys 1570, 1570C, 1571, 1575, 1580, 1597) [38].

The efficiency of substituting Sc for Yb, Er or Hf in Al-6%Mg alloys will now be analyzed.

The REEs Er and Yb form the intermetallics $Al_3Er$ and $Al_3Yb$ with $L1_2$ crystal lattice and form a continuous series of solid solutions with the $Al_3Sc$ phase. Therefore, solubility of Er and Yb in the $Al_3Sc$ phase can reach 100%. It should be noted that the atomic radii of Er (178 pm) and Yb (194 pm) are different from the atomic radius of Sc (162 pm). The use of Yb and Er as AEs increases the difference between the crystal lattice parameters of $Al_3(Sc_xMe_{1-x})$ and aluminum, and, consequently, increases the degree of noncoherence of the crystalline lattice of secondary particles $Al_3(Sc_xMe_{1-x})$.

The distribution coefficient of Yb and Er in aluminum is K << 1 [35]. This means that Yb and Er are horophilic elements in Al and should be concentrated predominantly along the grain boundaries during crystallization. The grain boundaries of coarse-grained metals ($d \sim 50–150$ μm), however, have small length ($\delta/d \sim 6·10^{-4}$ %, where $\delta = 2b$ is the grain boundary width, $b = 0.286$ nm is the Burgers vector of Al). In addition, β-phase particles are located along the grain boundaries of aluminum alloys. This precludes 0.016 at.% Yb and Er from being placed in the grain boundaries of a coarse-grained metal, and a significant part of the Yb and Er concentration is located in the grain volume.

It is important to emphasize that Yb and Er have low solubility in the aluminum crystal lattice [39, 40]. This leads to the formation of a large number of primary particles at the crystallization stage (Fig. 2) and, consequently, to a decrease in the concentration of Yb and Er in the solid solution. Low SER of Alloys #1-Yb and #2-Er are an indirect indication of a low concentration of Yb and Er in the solid solution. Table 2 demonstrates that the experimental values of SER ($\rho_{exp}$) of Alloys #1-Yb and #2-Er are ~0.2 μΩ·cm lower than their theoretical values $\rho_{th}$ that are calculated under the assumption of the contributions additivity of AEs which are completely in the solid solution. The procedure of calculating the theoretical value $\rho_{th}$ is described in detail in [41].

In our opinion, the noncoherent nature and small volume fraction of the precipitated $Al_3(Sc_xYb_{1-x})$ and $Al_3(Sc_xEr_{1-x})$ particles causes lower hardness and strength of the alloys in the annealed state (Figs. 3, 5). In addition, dislocations cannot cut noncoherent $Al_3(Sc_xYb_{1-x})$ and $Al_3(Sc_xEr_{1-x})$ particles, causing disclination defects to be formed on them and, consequently, accelerated cavitation destruction of Alloys #1-Yb and #2-Er. This precludes high elongation to failure during superplastic deformation of Alloys #1-Yb and #2-Er from being achieved (Table 3). If Yb and Er are to be efficient as AEs for Al-Mg-Sc-Zr alloys, it is necessary to use strong melt overheating and ultra-rapid crystallization methods. This will ensure a high concentration of Yb and Er in the aluminum crystal lattice and, consequently, increase their positive effect on Al-Mg-Sc-Zr alloys properties. It should be noted, however, that the efficiency of such approaches for producing large workpieces at steelworks needs to be discussed and assessed for economic feasibility.

Now the efficiency of substitution (0.1 wt.%) of Sc or Zr with 0.1 wt.% Hf will be analyzed.

Hf is known to dissolve well in the $Al_3Sc$ phase, forming solid substitutional solutions $Al_3(Sc_{1-x}Hf_x)$ with the $L1_2$ structure [17, 42]. In particular, Hf can replace more than 30 at.% Sc in $Al_3Sc$ particles [42]. The distribution coefficient of Hf in Al is $K_{Hf(Al)}$ = 4.6-6.01 [35] and, therefore, hafnium is concentrated predominantly in the aluminum grain volume. This suggests that, when heated, $Al_3(Sc_xHf_{1-x})$ particles predominantly precipitate throughout the volume of grains while $Al_3(Sc_{0.5}Zr_{0.5})$ particles precipitate along the grain boundaries of Alloy #3-Hf. Data on Hf solubility in Al are contradictory, but it is generally believed that Hf dissolves quite well in aluminum: ~ 1 wt.% at 660 °C and ~ 0.25–0.3% at 400 °C [43]. The increased solubility of Hf in Al has a controversial impact on the properties of the UFG Al-6%Mg-Sc-Zr alloy. On the one hand, the high solubility of Hf in Al can help to avoid precipitation of primary particles at the alloy crystallization stage and, consequently, to minimize the intensity of cavitation fracture of a UFG alloy. On the other hand, the high solubility of Hf in Al prevents the Al-Hf solid solution from decomposing completely at high annealing temperatures. This, in turn, does not allow the maximum volume fraction of $Al_3(Sc_xHf_{1-x})$ particles to be achieved and, consequently, high strength and thermal stability of a UFG alloy microstructure cannot be ensured.

The findings come into some contradiction with the data in [17] where it was demonstrated that the fine-grained Al-0.16%Sc-0.08%Zr-0.08%Hf alloy has a noticeably higher hardness and recrystallization temperature than the Al-0.17%Sc and Al-0.17%Sc-0.15%Hf alloys. The properties of the Al-0.16%Sc-0.08%Zr-0.08%Hf alloy were close to those of the Al-0.17%Sc-0.16%Zr alloy, which manifested the highest strength, hardness, and onset recrystallization temperature [17]. In our opinion, this contradiction can be due to the influence of magnesium on the nature and intensity of particles precipitation in the Al-Sc-Hf and Al-Sc-Zr-Hf alloys. Magnesium ($K_{Mg(Al)}$ = 0.304-0.6 [35]) decreases the aluminum grain boundary diffusion factor [11] and, consequently, slows down precipitation of $Al_3(Sc_xZr_{1-x})$ particles along grain boundaries. This precludes synergies of the contribution of $Al_3(Sc_xHf_{1-x})$ particles precipitating throughout the grain volume and of the contribution of $Al_3(Sc_xZr_{1-x})$ particles precipitating at grain boundaries.

**5. Conclusions**

1. The maximum elongation to failure of Al-6%Mg-0.12%Sc-0.10%Zr-0.10%Yb (Alloy #1-Yb) and Al-6%Mg-0.12%Sc-0.10%Zr-0.10%Er (Alloy #2-Er) ultrafine-grained alloys is observed at lower deformation temperatures than for Al-6%Mg-0.20%Zr-0.12%Sc (Alloy #4-Zr) and Al-6%Mg-0.22%Sc-0.10%Zr (Alloy #5-Sc). The maximum elongation to failure of Al-6%Mg-0.12%Sc-0.10%Zr-0.10%Hf (Alloy #3-Hf) is over 1,000%, exceeding the superplastic properties of Alloy #4-Zr and Alloy #5-Sc with a high content of alloying elements (in at.%). Alloy #1-Yb manifests good superplasticity characteristics ($\delta$ = 910%) at low temperatures (400 ºC). During superplastic deformation, competition is observed between strain-induced (dynamic) grain growth and dynamic recrystallization. The fracture of alloys under superplasticity occurs when pores are formed and grow rapidly.

2. Partially replacing Sc or Zr with a similar mass of Yb or Er is not effective due to the low solubility of these metals in the aluminum crystal lattice. This causes large primary $Al_3Yb$ and $Al_3Er$ particles to form at the crystallization stage, which further cause cavitation fracture of these alloys at elevated temperatures and higher strain rates. Precipitation of primary particles decreases the concentration of Yb and Er in the solid solution and, consequently, decreases the volume fraction of secondary $Al_3Yb$ or $Al_3Er$ nanoparticles that precipitate during heating. This makes it impossible to achieve high strength and hardness of UFG alloys or a high stability of the non-equilibrium microstructure of UFG Alloys #1-Yb and #2-Hf.

3. Partial replacement of Sc and Zr with Hf content of a similar mass proved to be the most effective. The UFG alloy is almost completely free of large primary particles, ensuring high values of elongation to failure. The effectiveness of using Hf as a partial replacement for Sc or Zr is limited by its high solubility in the aluminum crystal lattice, which decreases the volume fraction of precipitating secondary $Al_3(Sc_xHf_{1-x})$ particles and insufficient strength of annealed UFG Al-6%Mg alloys.

**Acknowledgments:** This research was funded by Russian Science Foundation, grant number 22-13-00149. The authors thank V.V. Zakharov (VILS, JSC; Moscow) for recommendations on the chemical composition and casting modes of aluminum alloys.

**Conflict of interest**. The authors declare that they have no conflict of interest.

**Credit Author Statement**: V.N. Chuvil'deev: Project administration, Funding acquisition, Methodology, Conceptualization, Supervision, Formal analysis, Writing – review & editing; M.Yu. Gryaznov & S.V. Shotin: Investigation (Tension test); A.V. Nokhrin & O.E. Pirozhnikova: Formal analysis, Writing – original draft, Writing – review & editing, Data curation; G.S. Nagicheva, C.V. Likhnitskii & I.S. Shadrina: Investigation (SEM, Metallography, Hardness, EDS); M.K. Chegurov: Investigation (Fractography); V.I. Kopylov: Investigation (ECAP); A.A. Bobrov: Investigation (Casting).